\documentclass[sigconf, nonacm]{acmart}
\usepackage{multirow}
\usepackage{float}
\usepackage{listings}
\usepackage{enumitem}

\newcommand\vldbdoi{XX.XX/XXX.XX}
\newcommand\vldbpages{XXX-XXX}
\newcommand\vldbvolume{14}
\newcommand\vldbissue{1}
\newcommand\vldbyear{2020}
\newcommand\vldbauthors{\authors}
\newcommand\vldbtitle{\shorttitle} 
\newcommand\vldbavailabilityurl{URL_TO_YOUR_ARTIFACTS}
\newcommand\vldbpagestyle{plain} 

\begin{document}
\title{DAgent: A Relational Database-Driven Data Analysis Report Generation Agent}

\settopmatter{authorsperrow=4}

\author{Wenyi Xu}
\affiliation{%
  \institution{Zhejiang University}
  \country{}
  }
\email{xuwenyi@zju.edu.cn}

\author{Yuren Mao}
\affiliation{%
  \institution{Zhejiang University}
  \country{}
  }
\email{yuren.mao@zju.edu.cn}

\author{Xiaolu Zhang}
\affiliation{%
 \institution{Ant Group}
 \country{}
 }
\email{yueyin.zxl@antfin.com}

\author{Chao Zhang}
\affiliation{%
  \institution{Zhejiang University}
  \country{}
  }
\email{zjuzhangchao@zju.edu.cn}

\author{Xuemei Dong}
\affiliation{%
  \institution{Zhejiang University}
  \country{}
  }
\email{xm.dong@zju.edu.cn}

\author{Mengfei Zhang}
\affiliation{%
  \institution{Zhejiang University}
  \country{}
  }
\email{zmengfei@zju.edu.cn}

\author{Yunjun Gao}
\affiliation{%
  \institution{Zhejiang University}
  \country{}
  }
\email{gaoyj@zju.edu.cn}

\begin{abstract}
Relational database-driven data analysis (RDB-DA) report generation, which aims to generate data analysis reports after querying relational databases, has been widely applied in fields such as finance and healthcare. Typically, these tasks are manually completed by data scientists, making the process very labor-intensive and showing a clear need for automation. Although existing methods (e.g., Table QA or Text-to-SQL) have been proposed to reduce human dependency, they cannot handle complex analytical tasks that require multi-step reasoning, cross-table associations, and synthesizing insights into reports. Moreover, there is no dataset available for developing automatic RDB-DA report generation. To fill this gap, this paper proposes an LLM agent system for RDB-DA report generation tasks, dubbed DAgent; moreover, we construct a benchmark for automatic data analysis report generation, which includes a new dataset DA-Dataset and evaluation metrics. 
DAgent integrates planning, tools, and memory modules to decompose natural language questions into logically independent sub-queries, accurately retrieve key information from relational databases, and generate analytical reports that meet the requirements of completeness, correctness, and conciseness through multi-step reasoning and effective data integration. Experimental analysis on the DA-Dataset demonstrates that DAgent's superiority in retrieval performance and analysis report generation quality, showcasing its strong potential for tackling complex database analysis report generation tasks.
\end{abstract}

\maketitle

\pagestyle{\vldbpagestyle}
\begingroup\small\noindent\raggedright\textbf{PVLDB Reference Format:}\\
\vldbauthors. \vldbtitle. PVLDB, \vldbvolume(\vldbissue): \vldbpages, \vldbyear.\\
\href{https://doi.org/\vldbdoi}{doi:\vldbdoi}
\endgroup
\begingroup
\renewcommand\thefootnote{}\footnote{\noindent
This work is licensed under the Creative Commons BY-NC-ND 4.0 International License. Visit \url{https://creativecommons.org/licenses/by-nc-nd/4.0/} to view a copy of this license. For any use beyond those covered by this license, obtain permission by emailing \href{mailto:info@vldb.org}{info@vldb.org}. Copyright is held by the owner/author(s). Publication rights licensed to the VLDB Endowment. \\
\raggedright Proceedings of the VLDB Endowment, Vol. \vldbvolume, No. \vldbissue\ %
ISSN 2150-8097. \\
\href{https://doi.org/\vldbdoi}{doi:\vldbdoi} \\
}\addtocounter{footnote}{-1}\endgroup

\ifdefempty{\vldbavailabilityurl}{}{
\vspace{.3cm}
\begingroup\small\noindent\raggedright\textbf{PVLDB Artifact Availability:}\\
The source code, data, and/or other artifacts have been made available at \url{https://github.com/ZJU-LLMs/DAgent}.
\endgroup
}


\section{Introduction}

Data analysis plays a crucial role in decision-making, which enables many significant discoveries such as economic trend prediction, financial risk assessment, and market behavior analysis. In enterprises, most of the data is still stored in relational databases. Analyzing this data requires professional skills, including writing complex SQL queries and analyzing execution results, and data scientists are often employed for this purpose. However, this expertise is unaffordable for many micro, small and medium-sized companies. Therefore, there is a pressing need to lower the barrier to this kind of relational database-driven data analysis (RDB-DA).

Although some methods, such as TableQA\cite{lu2024large} and Text-to-SQL \cite{hong2024next}, have been proposed to provide a user-friendly natural language interface for relational databases, these methods focus narrowly on lookup tasks—retrieving single values or simple aggregates. They do not support for more complex analytical workflows requiring cross-table reasoning, multi-step operations, or synthesizing insights into actionable reports, all of which are essential for RDB-DA. Thus, these methods can not provide a fundamental solution for RDB-DA. 
For example, answering a question such as "What are the sales trends across regions and how do they correlate with marketing spend?" might be beyond their capabilities, as they focus on single-query responses rather than integrated analysis.
To address this limitation, automatic data analysis report generation, which generates analytical reports based on natural language data analysis questions and the relational database, is a natural choice. However, it remains a relatively less-explored area.

There is not even a benchmark specifically for this task. Most existing datasets(e.g.,Spider\cite{2018arXiv180908887Y} and BIRD\cite{2023arXiv230503111L}) are mainly designed for retrieval tasks, usually aiming to locate specific values or direct results in a database. 
These datasets focus on fine-grained query tasks, with insufficient support for higher-level analytical tasks, especially in scenarios that require reasoning across multiple tables, integrating large-scale information, and generating comprehensive analytical reports.  
To fill this gap, we first introduce a benchmark for automatic data analysis report generation, which includes a new dataset, DA-Dataset, and evaluation metrics. This benchmark is specifically designed to evaluate methods for generating complete and accurate analytical reports from relational databases.

Then, we propose DAgent, an LLM-based agent system specifically designed to generate analytical reports from relational data-bases. 
It consists of three key modules: planning, tools, and memory. The planning module dynamically determines the execution path and selects appropriate tools based on the input question and the current execution context. The tools module includes specialized components such as problem decomposition tools, data retrieval tools, SQL rewriting tools, and answer generation tools. These components work together to analyze natural language questions, decompose them if necessary, then retrieve relevant data from relational databases, and synthesize the results into insightful analytical reports. The memory module records historical questions, corresponding execution paths, and intermediate results generated during the analytical process.

This LLM agent framework enables DAgent to address the inherent complexities of analytical tasks. By combining multi-step reasoning, cross-table associations, and data integration, DAgent effectively transforms questions into meaningful and logically coherent reports. 
Combined with the DA-Dataset, which offers a robust and realistic evaluation benchmark for real-world database analysis, our approach highlights the system's strengths in achieving completeness, correctness, and conciseness. Experimental results demonstrate that DAgent consistently outperforms existing methods, showcasing its strong potential for practical database analysis applications.

In summary, the key contributions of this paper can be outlined as follows:
\begin{itemize}
    \item We introduce \textbf{DAgent}, an LLM agent tailored for relational database-driven analytical report generation. DAgent integrates planning, tools, and memory modules, enabling it to adaptively handle complex analytical tasks.
    \item We construct a new dataset, \textbf{DA-Dataset}, specifically designed for analytical report generation tasks. This dataset addresses the limitations of existing datasets by emphasizing multi-table reasoning, summarization from large-scale data, and analytical report synthesis.
    \item Through extensive experiments, we demonstrate that DAgent outperforms state-of-the-art methods in report generation, achieving superior performance in completeness, correctness, and conciseness, highlighting its practical applicability and robustness in database analysis scenarios.
\end{itemize}

\vspace{-10pt}

\section{Related work}

Relational database-driven analytical report generation is a complex task, involving retrieving information from structured relational databases, performing cross-table associations, conducting multi-step reasoning, and generating analytical reports. Although there has been extensive research in fields such as Table Question Answering (Table QA) and Text-to-SQL, there are still significant research gaps in generating comprehensive analytical reports. This section will review the main areas relevant to this research, including Table QA, Text-to-SQL, and related datasets.

\subsection{Table QA}

The goal of TableQA is to obtain relevant information from tables or generate answers based on natural language questions. Existing methods can be broadly categorized into four types: semantic parsing, generation, extraction, and retrieval-reader methods\cite{jin2022survey}.

Semantic parsing methods mainly transform questions into SQL query statements to retrieve data. SQLNet\cite{xu2017sqlnet} uses a sketch-based approach to improve the accuracy of SQL generation.  Following this, TypeSQL\cite{yu-etal-2018-typesql} leverages type information from external knowledge bases to better understand rare entities.
Generation methods directly generate answers from tables. FeTaQA \cite{2021arXiv210400369N} uses an end-to-end pre-trained model to encode questions and tables and generate free-form answers. Further, FinQANet \cite{2021arXiv210900122C} focuses on numerical reasoning over financial data, supporting complex arithmetic operations.

Extraction methods select relevant cells as answers. For instance, TAPAS \cite{herzig2020tapas} incorporates table structure into cell representations using row and column embeddings. Mate \cite{2021arXiv210904312E} uses a multi-view attention mechanism to model table structure efficiently.
The retrieval-reader method is widely applied in open-domain TableQA. TableRAG \cite{chen2024tablerag} combines schema retrieval and cell retrieval, significantly improving efficiency when dealing with complex large-scale tables. ERATTA \cite{roychowdhury2024eratta} introduces an extreme retrieval-enhanced generation framework specifically designed for enterprise-level tasks, which incorporates features such as user authentication, query routing, and modular SQL generation.

Although these methods have achieved remarkable results, they still face numerous challenges in aspects such as multi-step reasoning, cross-table association, and meeting the requirements of specific domains. 

\subsection{Text-to-SQL}

Text-to-SQL aims to convert natural language queries into executable SQL statements, enabling non-expert users to access data-bases \cite{deng-etal-2022-recent, 2024arXiv240715186S}. Early methods primarily relied on handcrafted rules, such as Parsing Tree and Grammar Rule approaches, but these lacked generalization capabilities.

To overcome this limitations, deep learning methods, such as seq2seq architecture, improved SQL parsing. IRNet\cite{2019arXiv190508205G} uses an encoder for the representation of questions and schema, while RAT-SQL\cite{2019arXiv191104942W}, LGESQL\cite{2021arXiv210601093C} and S$^2$SQL\cite{2022arXiv220306958H} utilize graph neural networks to capture the alignment relationships between questions and schemas, subsequently employing a decoder to generate SQL queries. 
Pre-trained language models such as BERT, T5, mT5, and LLaMA have significantly enhanced the performance of Text-to-SQL tasks. Researchers fine-tuned these models to harness their robust semantic and linguistic capabilities, enabling the generation of precise SQL queries, such as TaBERT\cite{2020arXiv200508314Y} and REGROUP\cite{2023arXiv230101067D}.

Recently, LLMs have shown strong generalization capabilities in Text-to-SQL, significantly improving SQL generation through zero-shot and few-shot in-context learning and fine-tuning. For example, DIN-SQL\cite{2023arXiv230411015P} adopts a decomposition-based approach combined with self-supervised learning to enhance SQL generation, while DAIL-SQL\cite{2023arXiv230815363G} improves few-shot selection strategies to enhance LLM generalization across different databases. Additionally, C3\cite{dong2023c3} uses Clear Prompting, Calibration with Hints, and Consistent Output to improve SQL stability, and ACT-SQL \cite{2023arXiv231017342Z} automatically selects optimal prompt examples to improve complex query interpretation.

\subsection{Related datasets}
Currently, existing related datasets are mainly designed for Text-to-SQL and TableQA tasks.These datasets can generally be classified into two categories: Text-to-SQL datasets and TableQA datasets.

Among Text-to-SQL datasets, they can be further subdivided into original datasets and post-annotated datasets\cite{hong2024next}. 
Original datasets include Spider\cite{2018Spider}, WikiSQL\cite{2017wikisql}, BIRD\cite{2023bird}, etc. Take the BIRD dataset as an example. It was released in May 2023 and contains 12,751 examples, 95 databases. On average, each database has 7.3 tables and 549K rows of data. It has the characteristics of cross-domain and knowledge-augmentation.
Post-annotated datasets include ADVETA\cite{adveta}, Spider-SS\&CG\cite{spider-ss-cg}, etc. ADVETA was released in December 2022. Based on datasets such as Spider, it evaluates the robustness of models through adversarial table perturbations. Spider-SS\&CG divides the instances in the Spider dataset into sub-instances to explore context-dependent SQL generation.

TableQA datasets can be categorized into three main types: Closed-Domain Datasets like WTQ\cite{wtq} and SQA\cite{sqa}, which focus on specific domains and limited table contexts; Open-Domain Datasets such as OTTQA\cite{ottqa} and NQ-tables\cite{nq-tables}, designed for broader generalization across various tables; and Hybrid Datasets like TAT-QA\cite{2021arXiv210507624Z} and FinQA\cite{2021arXiv210900122C}, which combine tables with surrounding text and require models to handle both structured and unstructured data.

Even though these existing datasets are not directly suitable for analytical report generation tasks, their construction processes, including data collection strategies, preprocessing techniques, and annotation methods, could potentially offer precious insights. These insights can be harnessed to develop datasets that are specifically tailored for report generation, filling the current gap in this area.

\section{Preliminaries}

In this section, we first formally define the problem of relational database-driven analytical report generation and then introduce the background of LLM agent system, the Low-Rank Adaptation technique and the tabular data retrieval.

\subsection{Problem Statement}

Relational database-driven analytical report generation aims to generate correct and complete reports for natural language queries based on information stored in relational databases. Automatic relational database-driven analytical report generation can significantly facilitate data analysis in scenarios that requiring cross-table data integration and in-depth reasoning, such as financial statement analysis  \cite{ravi2020docubot, guo2024automatic}. In this context, the task involves extracting key financial metrics like total assets, liabilities, equity, and performance indicators, followed by the generation of a detailed report that evaluates the company’s financial health. This commonly involves complex database queries, statistical analysis, and detailed interpretation of the data to provide meaningful insights for decision-making.

Given a relational database consisting of a collection of tables $\mathcal{D} = \{T_1, T_2, \dots, T_N\}$, where each table $T_i$ within the database consists of columns $\{C_{i1}, C_{i2}, \dots, C_{iM}\}$, the task of relational database-driven analytical report generation can be formally defined as follows: Given a query $q$ and a relational database $\mathcal{D}$, the objective is to generate an analytical report $R$ which should satisfies three criteria: completeness, correctness and conciseness.

Completeness describes whether the generated report $R$ fully covers all the information required to satisfy the query $q$. ensuring no critical data is omitted. Correctness measures whether the generated report $R$ accurately reflects the intent of the query $q$. It focuses on whether the report content aligns with the query requirements, avoiding errors or semantic deviations. Conciseness, on the other hand, measures whether 
$R$ avoids excessive or redundant information, focusing solely on data that is directly relevant to the query. This ensures the report is clear, focused, and easy to interpret.

Relational database-driven analytical report generation differs significantly from existing paradigms such as TableQA\cite{lu2024large} and Text-to-SQL\cite{hong2024next}. TableQA mainly focuses on retrieving specific answers from tables, while Text-to-SQL obtain results by transforming natural language questions into SQL queries. Both methods are designed to return specific values or sets of values but typically do not involve analyzing or synthesizing the results. In contrast, the goal of analytical report generation is to produce comprehensive reports that require multi-step reasoning and information integration, which are essential for addressing complex query scenarios.

\vspace*{-10pt}
\addvspace{-5pt}
\subsection{LLM Agent System}

The LLM Agent System is a framework designed to address complex tasks by integrating reasoning, planning and execution capabilities within large language models. These systems are categorized into single-agent and multi-agent frameworks. Single-agent systems operate independently, leveraging methods like Chain-of-Thought (CoT)\cite{2022arXiv220111903W} reasoning and Tree-of-Thought (ToT)\cite{2023arXiv230510601Y}, while multi-agent systems involve the collaboration of multiple agents to solve large-scale problems through communication, task sharing, and collective intelligence, thereby enhancing scalability.\cite{guo2024large}.

At its core, an LLM Agent System consists of three core modules: planning, tools and memory\cite{xi2023rise}. The planning module serves as the decision-making hub, formulating strategies and decomposing tasks into subtasks based on the requirements of the environment. It can operate with or without feedback, adapting its plans dynamically based on environmental or human input. The tools module acts as the operational backbone, providing specialized functionalities such as API integration, database querying, and external model invocation, which enable the agent to execute actions beyond the inherent capabilities of LLMs. The memory module stores essential context, including historical interactions, intermediate results, and execution paths, allowing the agent to recall past experiences and refine future actions. This module often employs hybrid memory structures, combining short-term and long-term memory to enhance the agent's ability to reason over extended periods and complex scenarios.
A notable implementation of the LLM agent system is the ReAct method \cite{yao2022react}, which integrates reasoning and action in a synergistic manner. By guiding the model to generate steps that alternate between "thinking" and "action", it effectively improves the performance of the model in open-ended tasks.

LLM Agent Systems find broad applications in scenarios that require complex reasoning and multi-step execution. They are widely used in social science for simulating human behavior, in natural science for automating experiments and data management, and in engineering for tasks such as code generation and robotics. These systems demonstrate remarkable versatility, enabling them to adapt to diverse domains and tasks, thereby pushing the boundaries of what autonomous agents can achieve.

\subsection{Low-Rank Adaptation}

To efficiently adapt LLMs to specific downstream tasks, fine-tuning is often necessary. However, full-parameter fine-tuning of LLMs is computationally expensive and requires substantial memory resources, making it impractical for many applications. To address this challenge, parameter-efficient fine-tuning (PEFT) techniques have been developed, which aim to adapt pre-trained models to new tasks by updating only a small subset of parameters. Among these techniques, Low-Rank Adaptation (LoRA)\cite{hu2021lora} has emerged as one of the most widely recognized and effective methods, achieving competitive performance across various tasks while significantly reducing resource requirements\cite{mao2025survey}.

LoRA modifies the weight update process by constraining the updates to a low-rank decomposition. Specifically, for a pre-trained weight matrix $W_0 \in \mathbb{R}^{d_1 \times d_2}$, LoRA represents the update $\Delta W$ as a low-rank product $\Delta W = B \cdot A$, where $B \in \mathbb{R}^{d_1 \times d_r}$ and $A \in \mathbb{R}^{d_r \times d_2}$, with $d_r \ll \min(d_1, d_2)$. The forward pass is then adjusted as:$h' = W_0 x + \Delta W x = W_0 x + B (A x),$ where $x$ is the input. During training, $ W_0$ remains frozen, and only the low-rank matrices $B$ and $A$ are updated, which significantly reduces the number of trainable parameters and overall computational costs. The trained LoRA weights can either be merged with the original pre-trained weights or used independently as a plug-and-play module during inference. This approach retains the integrity of the base model while providing flexibility and efficiency in fine-tuning.

\subsection{Tabular data retrieval}
In the process of generating analytical reports from relational databases, retrieving relevant information from tables is a critical step.  There are two main approaches to tabular data retrieval: direct encoding and retrieval, and SQL query-based retrieval. The direct encoding and retrieval method encodes the structure and content information of the table into high-dimensional vector representations and uses vector similarity calculation to achieve fast retrieval\cite{herzig2020tapas}. This method is suitable for large-scale table data processing scenarios and performs well especially in open-domain tasks. The SQL query-based retrieval method generates SQL query statements to transform natural language questions into structured database queries to locate relevant table content. A typical representative is the Text-to-SQL system\cite{hong2024next}, which supports structured querying and can handle certain multi-table association tasks and fine-grained queries. Both methods have their own advantages. Direct encoding is suitable for quickly locating relevant information, while the SQL query method is more flexible and can adapt to fine-grained data analysis scenarios.

\section{Methodology}

\subsection{Overall Framework}

\begin{figure*}[h!]
    \centering
    \includegraphics[width=1.0\textwidth]{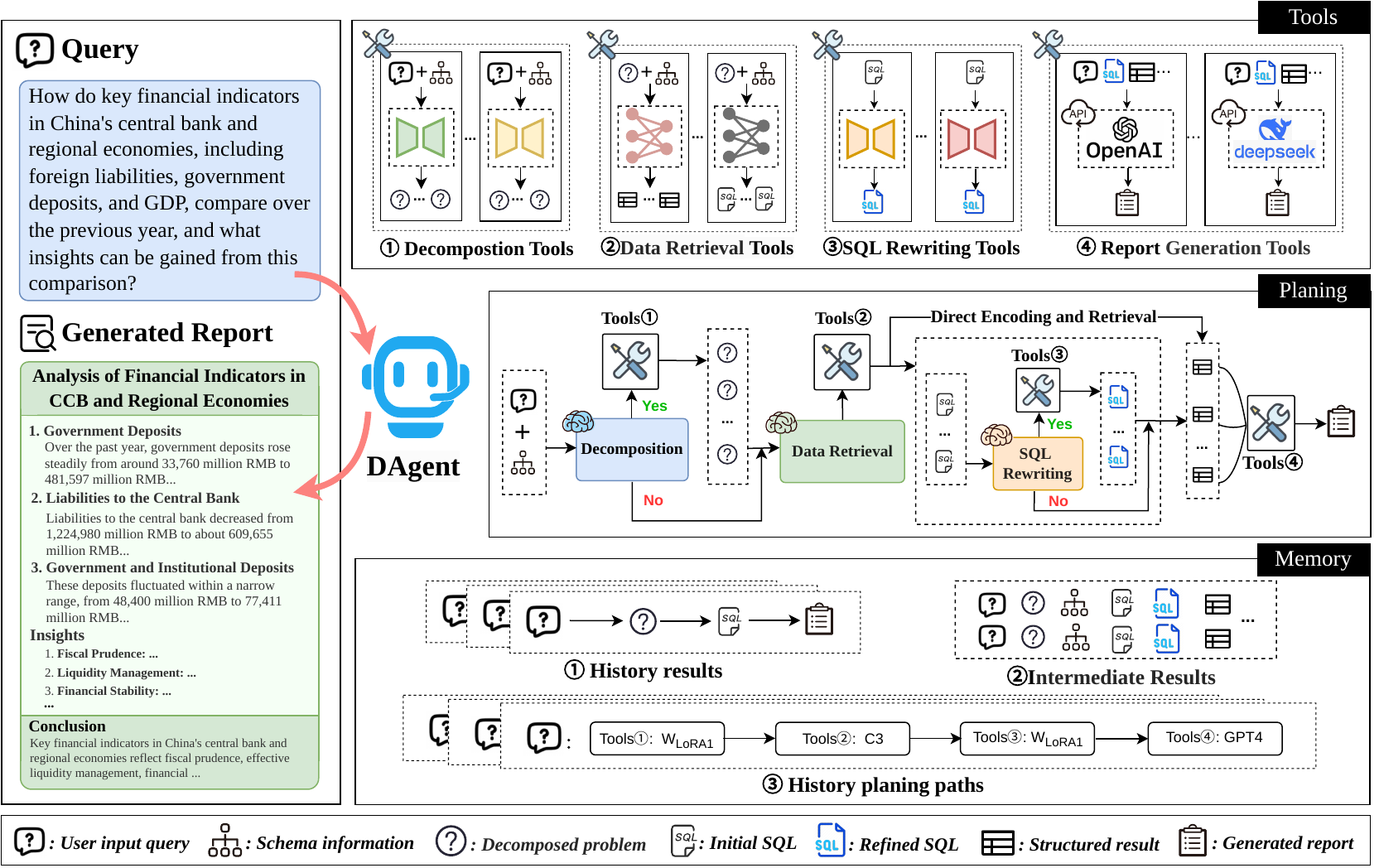} 
    \caption{System Overview of DAgent: An LLM Agent Framework for Relational Database-Driven Analytical Report Generation}
    \label{fig:workflow}
\end{figure*}

In this section, we introduced the overall design of DAgent, which is a modular agent system specifically designed for the task of relational database-driven analytical report generation. DAgent consisting of three main modules: Planning, Tools, and Memory, each module is responsible for distinct tasks and collaborates to accomplish database-driven data analysis. The Planning module is the "decision center" of DAgent, processes input queries, formulates execution strategies, and dynamically selects the appropriate tools. The Tools module serves as the execution unit, handling such as problem decomposition, tabular data retrieval, and report generation. The Memory module stores intermediate results, historical execution paths and results, providing long-term memory support to enhance report generation.

The workflow starts with a user's natural language question from the user. The Planning module first analyzes the input and determines wether the question needs decomposition. If the question is complex, the Planning module invokes the decomposition tools from tools module, breaking it down into sub-questions. These sub-questions are then processed by data retrieval tools, which retrieves relevant information from the database. Once the necessary data is gathered, the answer generation tool integrates the retrieved information with original question to produce a final analytical report. Throughout the process, the memory module records intermediate results, planning path and output, ensuring improved question efficiently and better contextual understanding for future tasks.

\subsection{Planning}

The Planning module is  responsible for formulating execution strategies and coordinating tool usage to process natural language questions accurately. It dynamically adjusts execution processes based on problem complexity, ensuring good collaboration between the tool module and the memory module. The Planning module handles question decomposition, selects appropriate data retrieval strategies, optimizes SQL queries, and coordinates analytical report generation.

When a natural language question $Q$ is received, the planning module first evaluates whether decomposition is required. The decision is based on factors such as question complexity, the number of distinct aspects involved, and whether multi-step reasoning. If decomposition is required, the planning module invokes the problem decomposition tool from the tools module to generate a set of sub-questions. If the question does not require decomposition, the question proceeds directly to the next stage to maintain execution efficiency and accuracy.

To illustrate this process, consider the question: \textit{"How do key financial indicators in China's central bank and regional economies, including foreign liabilities, government deposits, and GDP, compare over the previous year, and what insights can be gained from this comparison?"}. This question requires decomposition because it addresses multiple aspects, involving financial indicators, temporal comparison over the past year and analytical insights. Attempting to handle this question as a single task would lead to excessive complexity and a lack of clarity in the execution process. To address this, the planning module decomposes the question into sub-questions, such as:
\begin{itemize}
    \item \textit{"What are the government deposit trends over the past year?"}
    \item \textit{"How has GDP evolved over the past year in regional economies?"}
    \item \textit{"How have foreign liabilities to the central bank changed in the past year?"}
\end{itemize}

Once the question enters the next phase, the planning module determines how to retrieve relevant information from the database. The data retrieval process involves two key components: direct encoding and retrieval and SQL query-based retrieval. The direct encoding and retrieval tool is suited for fuzzy queries, while the Text-to-SQL tool is ideal for specific queries. Together, the retrieved information serves as the information source for report generation.

For direct encoding and retrieval, the planning module generates relevant keywords based on the question's intent. These keywords serve as search query to retrieve relevant database information. The specific retrieval process is managed within the tools, ensuring efficient indexing and similarity-based search to enhance retrieval accuracy. For SQL query-based retrieval, the planning module selects an appropriate Text-to-SQL model based on system resources. Given that Text-to-SQL models vary significantly in computational overhead and resource consumption, the module dynamically decides whether to utilize lightweight prompt-driven APIs for efficiency or deploy a locally fine-tuned language model for more precise SQL generation. This ensures optimal resource utilization while maintaining high query accuracy.

In the SQL query-based retrieval process, after the preliminary SQL query  $\text{SQL}_i$ is generated, the planning module will evaluate whether optimization is required to ensure that the query is both efficient and suitable for the subsequent answer-generation stage. Some SQL queries, when unoptimized, tend to yield excessively long results. This situation can bring about unnecessary computational overhead and also diminish the clarity of the retrieved data. To address these issues, the planning module will invoke the SQL-Rewrite tool in the tools module to analyze and optimize all SQL queries. The SQL-Rewrite tool will evaluate each query in terms of compactness and consistency with the analytical requirements of the question. If the query meets the criteria, the original query will be retained without modification. For queries that do not meet these standards, the SQL-Rewrite will optimize the query to generate a higher-quality SQL query $\text{SQL}_i'$.

For example, given the sub-question \textit{"What are the government deposit trends over the past year?"}, the Planning module first generates relevant keywords to guide the direct encoding and retrieval process. Simultaneously, it use Text-to-SQL method to generate an SQL query such as \textit{"SELECT GovernmentSavings FROM ed\_moneyauthoritybs;"}. If refinement is required, the Planning module invokes SQL optimization tool, producing an improved query like \textit{"SELECT Year, SUM(GovernmentSavings) AS TotalSavings FROM ed\_moneyauthoritybs GROUP BY Year;"}. This refined query ensures that the retrieved data aligns with the analytical needs of the original question.

After completing the data retrieval, the planning module invokes the report generation tool to generate the final analysis report. The planning module uses the currently stored content from the memory module, including direct encoding retrieval data, the optimized SQL query set $\{\text{SQL}_i'\}$, the corresponding execution results $\{R_i\}$, and the sub-question set $\{q_i\}$ generated during the problem decomposition stage. All this information, together with the original question $Q$, is passed to the tool module for use by the generation tool.

The purpose of this design is to ensure that the generation tool can have a comprehensive understanding of the context information. By combining the problem decomposition information and the optimized SQL results, it can generate an accurate and logically clear analysis report. At the same time, by directly invoking the tool module, the planning module decouples from the tool, leaving the specific implementation details to the tool module. This modular design improves the scalability of the system and the efficiency of task processing.

\subsection{Tools}
The tools module executes tasks based on the planning module's path. As a functional module called by the planning module, each tool serves a specific function to ensure efficient and accurate task completion. The tool module can be generally divided into four categories, including problem decomposition tools, tabular retrieval tools, SQL rewrite tools, and answer generation tools.

\subsubsection{Problem Decomposition Tools}

The core function of the problem decomposition tool is to break down a complex natural language question $Q$ into several logically independent sub-questions $\{q_i\}$. To meet the requirements of multi-domain tasks and improve the model's adaptability, the problem decomposition tool adopts the LoRA (Low-Rank Adaptation) technology for domain adaptation. The motivation for introducing LoRA lies in two main aspects: First, problem decomposition tasks usually involve multiple domains (such as finance, healthcare, etc.). By dynamically loading domain-specific LoRA weights, the tool can quickly adapt to the specific needs of different domains. Second, the low-rank update mechanism of LoRA does not require fine-tuning all the parameters of the large-scale model. This not only significantly reduces the training cost but also effectively protects the privacy of the original data. This design not only improves the efficiency of the tool but also enhances the model's security in sensitive scenarios.

In terms of technical implementation, the problem decomposition tool achieves dynamic domain adaptation by combining pre-trained weights and LoRA weights. The model formula is:
$M = W_{\text{base}}+\Delta W_{\text{LoRA}},$
where $W_{\text{base}}$ represents the pre-trained weights of the large language model, $\Delta W_{\text{LoRA}} = A\cdot B$ is the low-rank update weight of LoRA, $A\in\mathbb{R}^{d_r\times d_2}$ and $B\in\mathbb{R}^{d_1\times d_r}$ are trainable low-rank matrices, satisfying $d_r\ll\min(d_1, d_2)$. By loading specific $\Delta W_{\text{LoRA}}$ for each domain, the tool can efficiently adapt to multi-domain tasks without retraining the entire model.

Before decomposing the problem, the tool module selects the appropriate LoRA weights according to the domain to which the problem $Q$ belongs. The domain-selection process is completed by calculating the similarity between the problem embedding vector and the pre-defined domain keyword embedding vectors. These weights are then loaded and merged with the base model’s pre-trained weights, forming a domain-adapted model for task execution.
Once domain adaptation is complete, the module processes the input for problem decomposition,  including the natural language question $Q$ and the database schema information $\mathcal{S}$. These pieces of information are organized into a prompt template and input into the model to generate a set of sub-problems. The output of the model is a set of sub-problems $\{q_1,q_2,\dots,q_n\}$, where each sub-problem $q_i$ corresponds to an independent logical task.

\subsubsection{Data retrieval Tools}
Data retrieval tools are designed to extract relevant information from databases to support analytical report generation tasks, these tools are categorized into  the direct encoding retrieval tools and Text-to-SQL tools, each serving distinct roles in information retrieval.

The direct encoding and retrieval tool retrieves relevant tabular data through a two-step process: encoding and retrieval. 
In the encoding phase, the table schema and cell contents are transformed into high-dimensional vector representations. Formally, given a database $D$ with schema $S$, which includes tables $T = \{T_1,t_2...T_k\}$, and each table $T_i$ consisting of cell values $V_i=\{v_{ij} | j = 1,2, ... ,n\}$. 
The schema $ \mathcal{S} $ is encoded as $ E_{\mathcal{S}} $. For each cell $ v_{ij} $ in table $ T_i $, the model concatenates its table metadata, column metadata and the cell value then generates a contextualized embedding $ e_{v_{ij}} $. Cell embeddings are aggregated into matrices $ E_{V_i} $, ensuring semantic disambiguation through column-context fusion. In the retrieval phase, the tool performs a two-step search process. First, it compares the question embedding to the schema embeddings $E_s$ to identify the most relevant tables from the database. Once the relevant tables are identified, the system proceeds to the second step, where it uses the question to search within the selected tables and retrieve the most relevant cells.

The Text-to-SQL tools are designed to generate SQL queries from natural language inputs. The input to the Text-to-SQL tools consists of a set of sub-questions $\{q_i\}$ and a database schema $\mathcal{S}$, which includes structured information such as table names, column names, and data types.
The Text-to-SQL tool is implemented in two main ways in the DAgent framework: through external API calls and locally deployed models. For low-resource environments, we employ the C3 framework~\cite{dong2023c3}, which enhances SQL generation consistency through Clear Prompting and Calibration with Hints. In high-resource environments, we use the FinSQL framework~\cite{zhang2024finsql}, which leverages LoRA technology to efficiently fine-tune parameters for financial domain data, addressing performance challenges in cross-database migration and reducing dependency on external APIs. These two implementations complement each other, enabling DAgent to adapt to diverse task requirements while optimizing system performance and resource utilization.

The direct encoding and retrieval tool is suited for fuzzy queries and the Text-to-SQL tool, on the other hand, is ideal for specific queries. Together, they provide a comprehensive solution for both broad and detailed data retrieval needs.

\subsubsection{SQL Rewrite Tools}

The primary purpose of the SQL rewrite tools is to optimize the generated SQL queries, aiming to reduce the amount of data and enhance the relevance between the results and the problem. The initially generated SQL may cover excessive data, resulting in a large-scale retrieval result, which increases the complexity of subsequent processing. Especially when complex analysis is required, too much irrelevant data can significantly interfere with the quality of the generated results. Therefore, the SQL rewrite tool optimizes the SQL structure to ensure that the generated results are more compact and in line with the problem requirements, thus improving the overall efficiency of the system and the accuracy of the results.

In terms of technical implementation, the SQL rewrite tool also adopts the LoRA to achieve efficient model update and domain adaptation. It can be expressed by the formula: $M = W_{\text{base}}+\Delta W_{\text{SQL}}$,
where $W_{\text{base}}$ represents the pretrained model weights, and $\Delta W_{\text{SQL}}$ is the low-rank update weight for the SQL rewrite task. By loading LoRA weights of different domains, the SQL rewrite tool can quickly adapt to the query optimization requirements of specific domains without retraining the entire model. The inputs of the SQL rewrite tool include the initially generated query $\text{SQL}_i$, the database schema $\mathcal{S}$, and sub-question $\mathcal{Q_i}$. The output is the optimized SQL query $\text{SQL}_i'$. If the initial query $\text{SQL}_i$ meets the optimization conditions, it will be output as it is; otherwise, the tool will rewrite its logic and streamline the fields to generate a high quality optimized query $\text{SQL}_i'$. Through this design, the SQL rewrite ensures the efficiency of data retrieval and the precision of query results.

\subsubsection{Report Generation Tools}

The primary purpose of the Report Generation Tools is to transform the retrieved data and contextual information into a comprehensive, semantically accurate, and logically coherent analytical report $R$. These tools play a crucial role in bridging structured database outputs and human-readable insights, ensuring the generated report meets the analytical and decision-making requirements specified by the query $Q$.

The Report Generation Tools integrate the retrieved results $\{R_i\}$ with contextual information, including the original question $Q$, sub-questions $\{q_i\}$, and optimized SQL queries $\{\text{SQL}_i'\}$, to ensure the report fully captures the intent of the query. By leveraging advanced language models, such as LLaMA or ChatGPT, the tool produces a structured and well-articulated report that aligns with the query's analytical objectives. These inputs are organized into a prompt template that guides the language model to synthesize relevant and high-quality content. The report generation process can be formalized as:
$R = LLM(\text{Prompt}),$ where $LLM$ represents the large language model.

\subsection{Memory}

In the DAgent system, the core function of the Memory module is to provide efficient support for task execution. By storing key information, the system can leverage historical experience in questions, thereby improving the efficiency and quality of task processing. Specifically, the Memory module achieves this goal through three types of memory: historical user questions and results, intermediate content for the question, and historical question and planning path correspondence.

Firstly, the Memory module stores historical user questions and results. Whenever DAgent processes a question and generates a corresponding report, the system stores the user’s question and its execution results in memory. This allows the system to quickly generate responses to similar question without needing to re-execute the data retrieval and analysis process, instead directly leveraging historical records.

Secondly, the Memory module stores intermediate generation content for the current question. During the processing of each user question, DAgent goes through multiple steps, including problem decomposition, data retrieval, etc. The intermediate data and results generated at each step are recorded in the Memory. This information provides essential contextual support for final report generation.

Finally, the Memory module also stores historical question and planning path correspondence. Each user question is associated with an execution path generated by the planning module, which includes key strategies such as whether the problem needs decomposition, which tools to use, etc. These historical planning paths provide valuable experiential support for DAgent when processing similar questions, helping the system quickly determine the most appropriate execution strategy, thereby improving task execution efficiency.

\begin{figure*}[!h]
    \centering
    \includegraphics[width=1.0\textwidth]{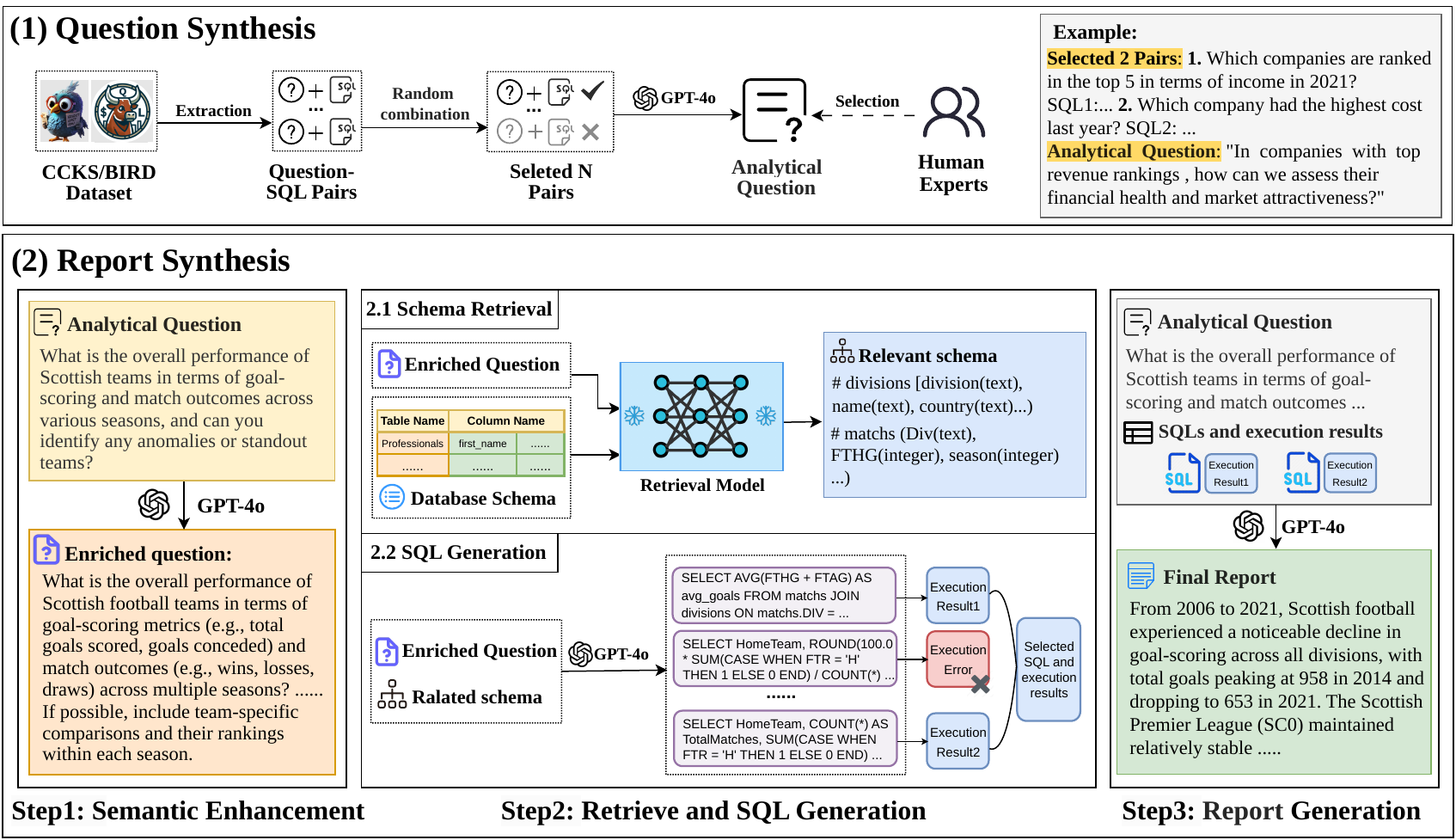}
    \caption{Step-by-Step Workflow for Constructing the DA-Dataset for Analytical Report Generation}
    \label{fig:aa} 
\end{figure*}

\section{DA-Dataset}

To comprehensively evaluate the performance of the DAgent framework in analytical report generation, we constructed a multi-domain dataset that focuses on analytical questions. While existing public datasets\cite{2023bird, spider-ss-cg} contain a wealth of basic questions and SQL queries, they are typically limited to retrieval of exact answers from database. They do not effectively simulate real-world data analysis problems, which often involve hierarchical structures and require multi-perspective analysis. To address this gap, we have designed a new dataset aimed at testing the DAgent's capabilities in information retrieval and report generation. 
The dataset contains a total of 735 entries, with 201 entries constructed from the HangSeng Financial dataset\cite{zhang2024finsql} (DA-CCKS) and 535 entries constructed from the BIRD dataset\cite{2023bird} (DA-BIRD). On average, each question requires retrieving information from 4.5 tables and 10.8 columns, reflecting the complexity and multi-faceted nature of the analytical tasks presented in the dataset.
As shown in Figure \ref{fig:aa}, the dataset construction involves two key steps: (1) query synthesis, and (2) report Synthesis. This design enables a more comprehensive evaluation of the system's ability to handle complex problems and provides a solid foundation for further optimizing the performance and applications of DAgent.

\subsection{Question Synthesis}


The dataset construction begins by designing analytical questions that emphasize summarization and hierarchical complexity, requiring the integration of diverse data to produce multi-layered answers. These questions simulate real-world scenarios demanding cross-perspective reasoning, aiming to evaluate the DAgent’s capabilities in handling complex tasks such as logical reasoning, data retrieval, and report synthesis.

In the implementation process, we first extracted all question-SQL pairs from the HangSeng Financial\cite{zhang2024finsql} and BIRD\cite{2023bird} datasets. These pairs served as the basis for generating analytical questions. Next, we randomly selected 10 question-SQL pairs and used them as input, feeding them to LLMs using a carefully designed prompt.  LLMs then generated a analytical question by combining the logic of these questions. To ensure the quality of the generated questions, all outputs were manually reviewed. Any questions that did not meet the required standards were modified or replaced to ensure semantic clarity and appropriate complexity. These generated questions typically integrate multiple sub-questions and demonstrate characteristics of summarization and multi-perspective analysis.


\subsection{Report Synthesis}
\vspace{2mm}
The second step generates an analytical report using the questions from the first step. It retrieves relevant data from the database and synthesizes it into a coherent natural language report. This process combines semantic enhancement, schema retrieval, SQL generation, and report synthesis to ensure accuracy and alignment with the original question's intent.

\vspace{0.8mm}
In implementation, we first use LLMs to semantically enhance the original question. The LLM analyzes the question, integrates contextual information, and optimizes its expression using the database schema. This step improves clarity and schema compatibility, facilitating subsequent schema retrieval.
Next, we use a cross-encoder to retrieve relevant database tables and columns by bidirectionally encoding the enhanced question and schema, calculating relevance scores. The LLM then generates SQL queries based on the refined question and schema. To ensure SQL quality, a dynamic filtering mechanism is applied: candidate SQLs are automatically executed, retaining those returning valid results, while manual review verifies semantic accuracy. The final SQLs strictly meet the original question’s semantic requirements.
After executing the filtered SQL queries, the corresponding structured data is retrieved from the database. These results are then converted into a natural language analysis report using the prompt template for natural language generation. The template is designed based on the question type and the structure of the SQL results, embedding key fields into fixed sentence patterns to produce a  semantically accurate report.

This step-by-step design facilitates efficient collaboration among semantic enhancement, schema retrieval, SQL generation, and natural language generation. Semantic enhancement ensures the clarity and interpretability of the question; schema retrieval and SQL query generation guarantee accurate data extraction; and template-based report generation elevates the quality of the natural language output. By combining the capabilities of the large model and the cross-encoder model, the system can effectively handle complex database queries, providing high-quality, logically coherent, and semantically consistent reports—demonstrating its enormous potential for practical applications.

\begin{figure}[!h]
    \centering
    \includegraphics[width=1.0\columnwidth]{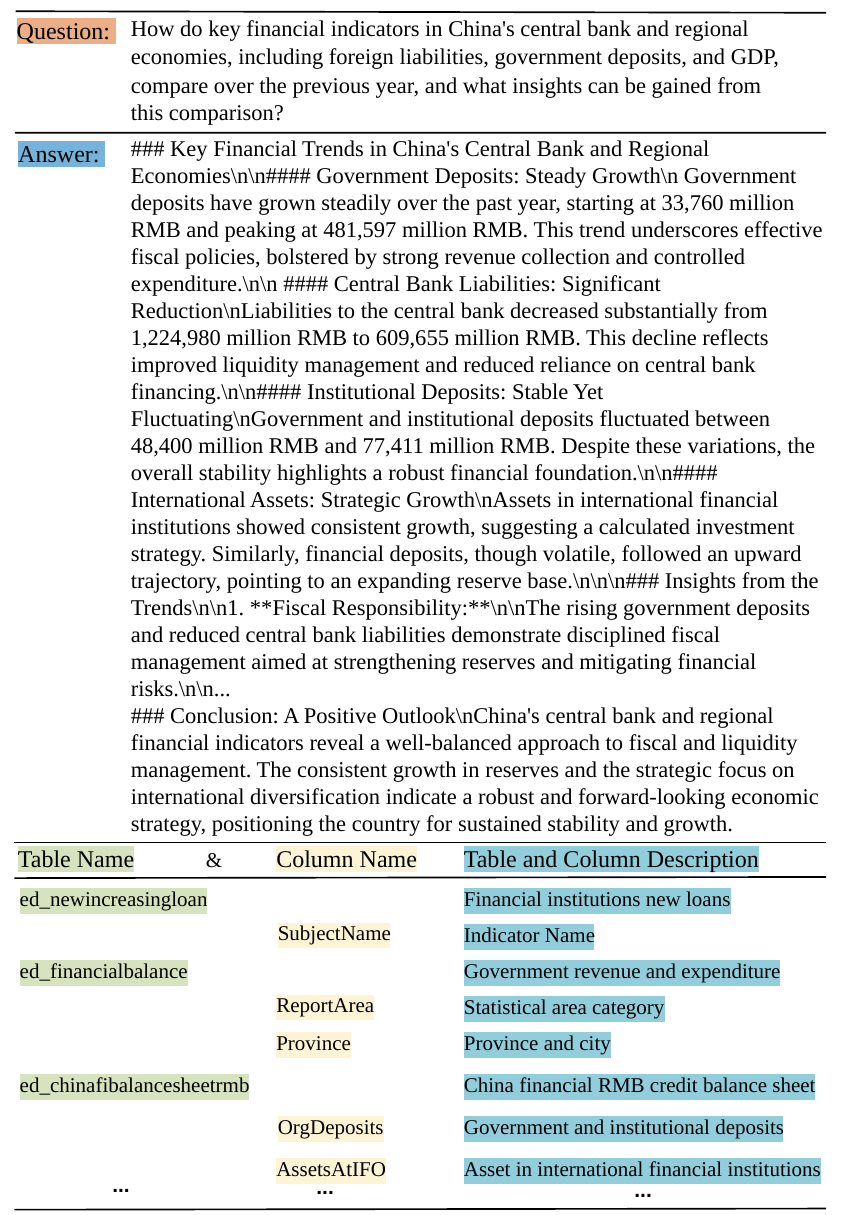}
    \caption{Example Entry from the DA-Dataset}
    \label{fig:example}
\end{figure}

\section{Experiments}

\subsection{Experimental settings}

\subsubsection{Methods for Comparison}
To comprehensively evaluate the performance of DAgent, we compare it with a series of methods, including baseline approaches, Text-to-SQL methods, and advanced table question-answering frameworks.  The comparison methods are categorized as follows:

\textbf{Baseline Methods}:  
(1) \textbf{LLM-Answer}: This method directly uses a large language model (LLM) to answer questions without relying on any intermediate steps, such as schema parsing or structured query execution. It entirely depends on the LLM's generative capabilities and pre-trained knowledge to infer and produce answers.
(2) \textbf{Schema-Answer}: This method incorporates database schema information, including metadata of tables and columns, into the input for the LLM. By combining this structured information, Schema-Answer enhances the model's understanding of the question and its contextual reasoning ability, generating more relevant answers.
(3) \textbf{RandomRowCol}: This method randomly selects content from table cells to generate answers without involving any semantic parsing, query generation, or schema understanding. It is a completely random and heuristic approach, serving as a baseline method to evaluate the effectiveness of more advanced methods.

\begin{table*}[!t]
    \centering
    \caption{Retrieval Performance Comparison of DAgent and Baseline Methods on DA-CCKS Dataset}
    \resizebox{\textwidth}{!}{
        \begin{tabular}{llccccccc}
        \toprule
        \multirow{2}{*}{Category} & \multirow{2}{*}{Method} & \multicolumn{3}{c}{Table Retrieval} & \multicolumn{3}{c}{Column Retrieval} & \multirow{2}{*}{Context Relevance} \\
        \cmidrule(lr){3-5} \cmidrule(lr){6-8}
         &  & Precision & Recall & F1-score & Precision & Recall & F1-score &  \\
        \midrule
        \multirow{3}{*}{Baseline} & LLM-Answer & --- & --- & --- & --- & --- & --- & --- \\
         & Schema-Answer & --- & --- & --- & --- & --- & --- & 2.95 \\
         & RandomRowCol & 24.06 & 15.28 & 17.97 & 6.70 & 3.00 & 4.00 & 1.79 \\
        \midrule
        \multirow{3}{*}{Text-to-SQL} & LLM-SQL & 24.86 & 12.46 & 15.04 & 8.10 & 6.90 & 6.30 & 3.75 \\
         & C3 & 33.14 & 26.17 & 28.54 & 12.72 & 11.92 & 11.58 & 4.83\\
         & FINSQL & 38.55 & 24.16 & 28.94 & 15.98 & 15.51 & 15.03 & 3.97 \\
        \midrule
        \multirow{2}{*}{TableQA} & TableRAG & 40.20 & 35.19 & 35.84 & 14.64 & 16.41 & 14.96 & 5.31 \\
         & ERATTA & 33.48 & 25.60 & 28.01 & 13.95 & 12.68 & 12.33 & 4.24 \\
        \midrule
        \textbf{Ours} & \textbf{DAgent} & \textbf{56.35} & \textbf{36.96} & \textbf{43.10} & \textbf{20.64} & \textbf{20.46} & \textbf{19.96} &  \textbf{6.87} \\
        \bottomrule
        \end{tabular}
    }
\label{tab:t1}
\end{table*}

\begin{table*}[]
    \centering
    \caption{Retrieval Performance Comparison of DAgent and Baseline Methods on DA-BIRD Dataset}
    \resizebox{\textwidth}{!}{
        \begin{tabular}{llccccccc}
        \toprule
        \multirow{2}{*}{Category} & \multirow{2}{*}{Method} & \multicolumn{3}{c}{Table Retrieval} & \multicolumn{3}{c}{Column Retrieval} & \multirow{2}{*}{Context Relevance} \\
        \cmidrule(lr){3-5} \cmidrule(lr){6-8}
         &  & Precision & Recall & F1-score & Precision & Recall & F1-score &  \\
        \midrule
        \multirow{3}{*}{Baseline} & LLM-Answer & --- & --- & --- & --- & --- & --- & --- \\
         & Schema-Answer & --- & --- & --- & --- & --- & --- & 3.25 \\
         & RandomRowCol & 57.50 & 58.32 & 55.49 & 36.02 & 17.75 & 22.72 & 1.40 \\
        \midrule
        \multirow{3}{*}{Text-to-SQL} & LLM-SQL & 54.21 & 58.28 & 54.64 & 42.85 & 48.95 & 43.89 & 4.57 \\
         & C3 & 60.88 & 53.24 & 55.57 & 48.95 & 53.65 & 49.83 & 5.53 \\
         & FINSQL & 65.65 & 66.25 & 63.53 & 51.36 & 58.82 & 52.97 & 5.63 \\
        \midrule
        \multirow{2}{*}{TableQA} & TableRAG & 65.59 & \textbf{88.97} & 72.77 & 51.15 & 61.26 & 53.89 & 5.76 \\
         & ERATTA & 61.37 & 59.51 & 59.12 & 46.31 & 58.58 & 51.41 & 5.16 \\
        \midrule
        \textbf{Ours} & \textbf{DAgent} & \textbf{81.23} & 88.82 & \textbf{82.23} & \textbf{69.39} & \textbf{77.53} & \textbf{71.21} & \textbf{7.15} \\
        \bottomrule
        \end{tabular}
    }
\label{tab:t2}
\end{table*}

\textbf{Text-to-SQL Methods}:  
(1) \textbf{LLM-SQL}: LLM-SQL generates SQL queries directly using large language models, transforming natural language questions into structured queries. It processes the input question along with database schema, including table and column metadata, to produce executable SQL. However, this approach does not deeply optimize for database-specific structures or semantics, making it more suitable for straightforward query scenarios rather than complex analytical tasks. 
(2) \textbf{C3}: C3\cite{dong2023c3} is a zero-shot Text-to-SQL method based on ChatGPT, comprising three core components: Clear Prompting, Calibration with Hints, and Consistent Output. By optimizing prompt design, C3 ensures clear semantic input; through bias calibration strategies, it avoids unnecessary columns or incorrect keywords; and by applying execution-based consistency checks, it selects the optimal query, ensuring the generated SQL aligns with the question's requirements.
(3) \textbf{FinSQL}: FinSQL\cite{zhang2024finsql} is a model-agnostic Text-to-SQL framework designed for financial databases, addressing challenges like wide-table schemas, data privacy constraints, and complex analytical requirements. It integrates hybrid data augmentation for prompt engineering, parameter-efficient LoRA fine-tuning, and SQL consistency calibration to optimize cross-database generalization. The framework achieves state-of-the-art performance in financial scenarios while reducing reliance on external computational resources.

\begin{table*}[]
    \centering
    \caption{Performance Comparison of Report Generation Quality}
    \resizebox{\textwidth}{!}{
        \begin{tabular}{llcccccc}
        \toprule
        \multirow{2}{*}{Category} & \multirow{2}{*}{Method} & \multicolumn{3}{c}{DA-CCKS} & \multicolumn{3}{c}{DA-BIRD} \\
        \cmidrule(lr){3-5} \cmidrule(lr){6-8}
         &  & Accuracy & Report Relevance & Average & Accuracy & Report Relevance & Average \\
        \midrule
        \multirow{3}{*}{Baseline} & LLM-Answer & 1.30 & 4.81 & 3.06 & 1.78 & 4.96 & 3.37 \\
         & Schema-Answer & 3.05 & 6.29 & 4.67 & 3.31 & 6.57 & 4.94 \\
         & RandomRowCol & 3.57 & 6.38 & 4.97 & 2.10 & 6.35 & 4.23 \\
        \midrule
        \multirow{3}{*}{Text-to-SQL} & LLM-SQL & 3.28 & 6.31 & 4.80 & 4.16 & 6.40 & 5.28 \\
         & C3 & 3.82 & 6.57 & 5.20 & 4.76 & 6.44 & 5.60 \\
         & FINSQL & 4.61 & 6.35 & 5.48 & 5.31 & 6.55 & 5.93 \\
        \midrule
        \multirow{2}{*}{TableQA} & TableRAG & 5.47 & 6.77 & 6.12 & 5.83 & 6.99 & 6.41 \\
         & ERATTA & 5.55 & 6.43 & 6.00 & 3.91 & 6.31 & 5.11 \\
        \midrule
        \textbf{Ours} & \textbf{DAgent} & \textbf{6.27} & \textbf{6.96} & \textbf{6.62} & \textbf{6.43} & \textbf{7.01} & \textbf{6.72} \\
        \bottomrule
        \end{tabular}
    }
\label{tab:t3}
\end{table*}

\textbf{Table QA Methods}:  
(1) \textbf{TableRAG}: TableRAG\cite{chen2024tablerag} is a retrieval-augmented generation framework that combines schema retrieval and cell retrieval to efficiently extract critical information from tables for answering questions. Schema retrieval identifies important columns and their data types, while cell retrieval pinpoints key values related to the query. By optimizing prompts and encoding cells independently, it enhances efficiency on complex/large tables, excelling in table understanding tasks.(2) \textbf{ERATTA}: ERATTA\cite{roychowdhury2024eratta} introduces an extreme retrieval-augmented generation approach tailored for enterprise-level table QA tasks. Its framework consists of modules for user authentication, query routing, SQL generation, and answer generation, enabling rapid information retrieval and answering from dynamic tables. ERATTA incorporates non-LLM hallucination detection metrics and employs a serialized task decomposition strategy, effectively handling complex multi-table queries, making it suitable for large-scale enterprise data applications.

\subsubsection{Dataset}

In our experiments, we use the DA-Dataset, specifically constructed to evaluate DAgent's performance in relational database-driven analytical report generation. The DA-Dataset is constructed with a focus on multi-perspective analytical reasoning, cross-table data retrieval, and the synthesis of comprehensive reports. It is divided into two subsets: DA-CCKS and DA-BIRD. DA-CCKS is derived from the HangSeng Financial text-to-SQL dataset and emphasizes financial analysis tasks, featuring questions and data structures tailored to real-world financial scenarios such as asset evaluation and risk management. DA-BIRD, on the other hand, is constructed from the BIRD dataset and encompasses a broader set of analytical tasks, with hierarchical and multi-perspective questions spanning diverse domains. On average, each question requires retrieving information from 4.5 tables and 10.8 columns, reflecting the complexity and multi-faceted nature of the analytical tasks presented in the dataset.

\subsubsection{Evaluation Metrics}
To evaluate DAgent comprehensively, we employ two categories of metrics: retrieval performance metrics and report generation quality metrics. These metrics are designed to assess the system’s ability to retrieve relevant information from relational databases and generate logically coherent analytical reports.

For retrieval performance, the evaluation measures how effectively the system covers the necessary data required to generate a detailed report. This is achieved by quantifying recall, which indicates the proportion of all relevant tables and columns that are successfully retrieved, as well as precision, which reflects the accuracy of the retrieval process by determining the proportion of correctly retrieved items among all the items fetched. In addition to these structural metrics, we assess context relevance to evaluate the quality and alignment of the content within the retrieved columns, following the advanced RAGAS\cite{es2023ragas} method used in RAG for evaluating the quality of retrieved content. 
 
\vspace{0.17mm}
For report generation quality, two key metrics are used: Accuracy and Report Relevance. Accuracy assesses whether the generated report correctly addresses the query requirements and reflects the intended semantic meaning. For Report Relevance, we reference the method RAGAS\cite{es2023ragas}, it evaluates whether the generated report is directly relevant to the query and avoids redundancy or incomplete information. In this approach, multiple questions are generated based on the provided report. The semantic similarity between these generated questions and the original query is then computed. The statistical average similarity score of these generated questions relative to the original query is used to evaluate the relevance of the generated report.

\begin{figure}[!h]
    \centering
    \includegraphics[width=1.0\columnwidth]{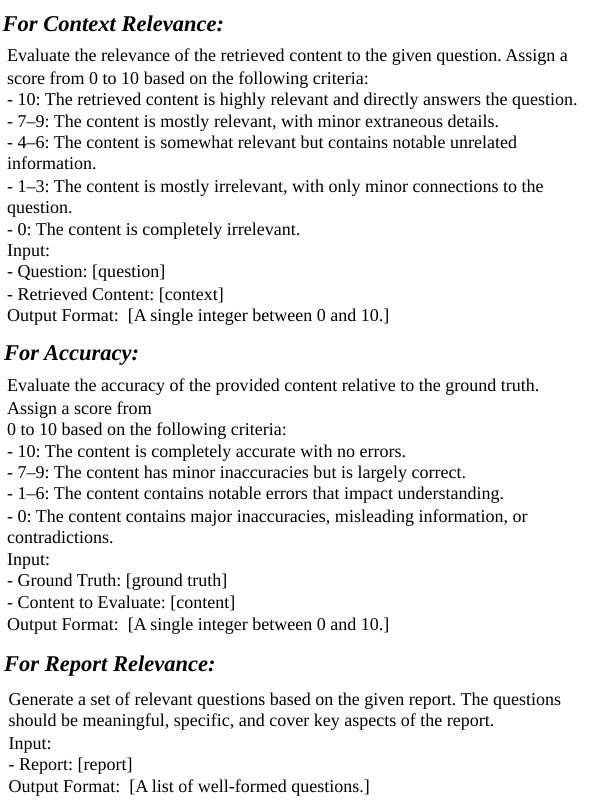}
    \caption{Evaluation Prompt Template: Retrieval Performance Metrics \& Report Quality Metrics}
    \label{fig:prompt}
\end{figure}

\vspace{0.17mm}
These metrics are evaluated through prompts provided to large language models. In our experiments, we use DeepSeek-V2.5 as the evaluation model. DeepSeek-V2.5 is prompted to assess each metric on a scale from 0 to 10. For each of the evaluation metrics, the relevant prompts are designed in figure \ref{fig:prompt}.

\begin{figure*}[!t]
    \centering
    \includegraphics[width=1.0\textwidth]{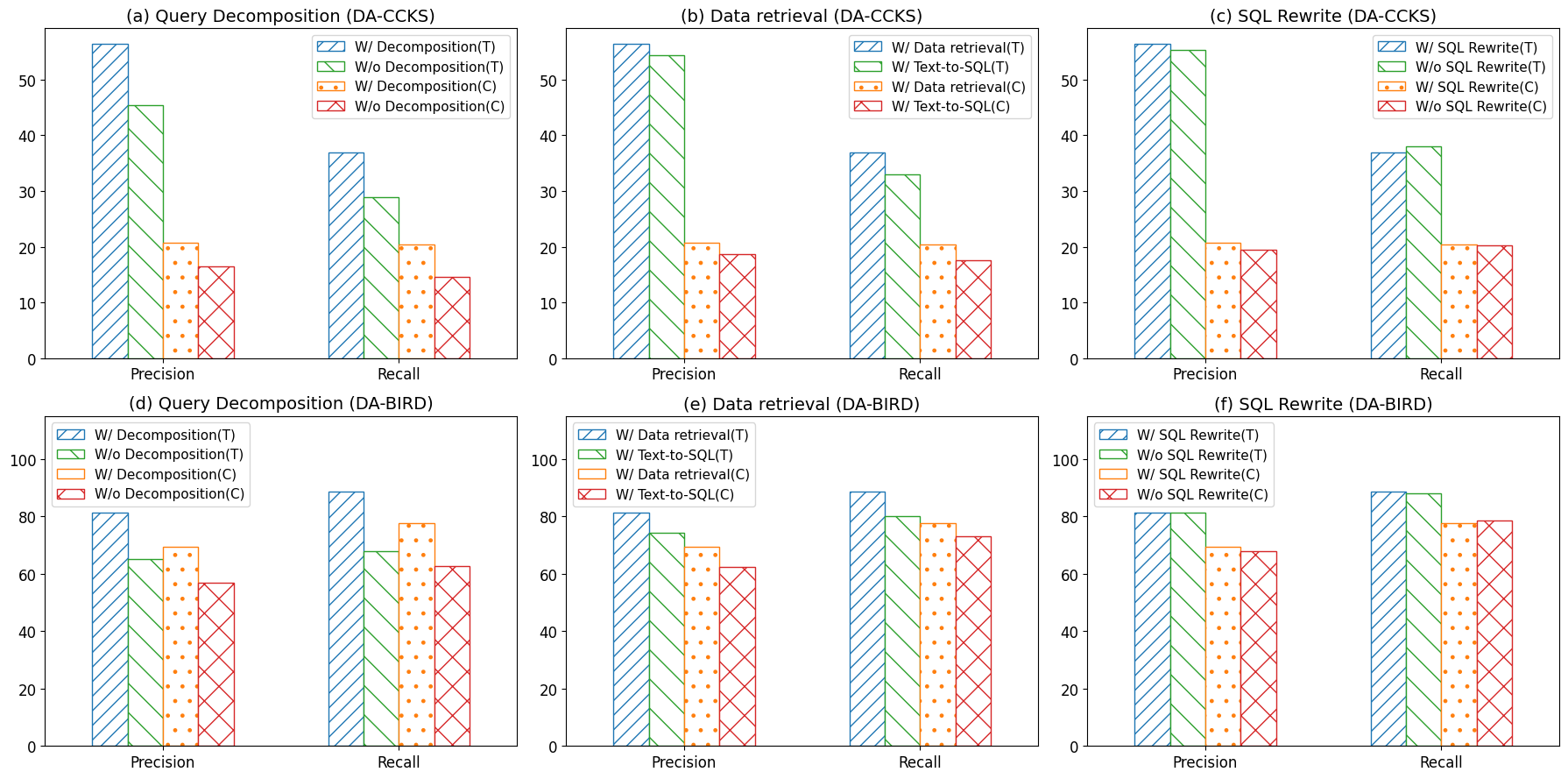}
    \caption{Ablation study results comparing the effects of different planning strategies in the DAgent system.}
    \label{fig:ab} 
\end{figure*}

\subsubsection{Implementation Details}

The DAgent system is implemented using LangChain as the primary framework, enabling seamless integration of planning, tools, and memory modules. The base model for LoRA is LLaMA2-13B. All LoRA-enabled modules, including those for problem decomposition and SQL Rewrite tools, are trained and inferred with a rank of $r=32$ and a learning rate of $1 \times 10^{-4}$.  The direct encoding and retrieval tool uses the text-embedding-ada-002 encoder. The report generation utilizes DeepSeek v2.5. Experiments are conducted on a single NVIDIA A40 GPU with 46GB memory.

\subsection{Overall Performance}
The experimental results are shown in Table \ref{tab:t1}, Table \ref{tab:t2}, and Table \ref{tab:t3} comprehensively demonstrate DAgent’s performance across retrieval and report generation tasks. DAgent achieves significant improvements over baseline methods in retrieval accuracy (Recall, Precision, and Context Relevance) and report quality (Accuracy and Report Relevance), demonstrating its effectiveness in handling complex analytical questions.

\textbf{Retrieval Performance Analysis}
In the retrieval performance analysis, we focused on evaluating DAgent's performance including table retrieval, column retrieval, and context relevance. As shown in Table \ref{tab:t1} and Table \ref{tab:t2}, the experimental results show that DAgent outperforms existing baseline methods in all these aspects, particularly in handling complex retrieval tasks, such as DA-CCKS.

DAgent demonstrates a significant advantage in balancing recall and precision. On the DA-CCKS dataset, DAgent achieves a Recall of 36.96, a Precision of 56.35, and an F1-score of 43.10, clearly outperforming all baseline methods. Similarly, on the DA-BIRD dataset, DAgent achieves a Precision of 81.23, a Recall of 88.82, and an F1-score of 82.23, substantially surpassing the best baseline method, TableRAG, which has an F1-score of 72.77. 
The experimental results confirm that DAgent effectively reduces the retrieval of irrelevant tables and columns while ensuring comprehensive recall. Thanks to the synergy of multi-step reasoning and multiple retrieval strategies, DAgent is able to efficiently pinpoint key information from multiple relevant tables in database. 
In terms of retrieval context quality, DAgent also excels. Under the unified evaluation template, DAgent achieves a context relevance score of 6.87 on the DA-CCKS dataset and 7.15 on the DA-BIRD dataset. This indicates that DAgent not only performs accurately in table and column matching but also retrieves highly relevant content with great precision.

\vspace{0.3mm}
\textbf{Generation Quality Analysis}
The generation quality analysis focuses primarily on the accuracy and relevance of the reports. By comparing DAgent’s performance with other methods in Table \ref{tab:t3}, we demonstrate its superiority in analytical report generation tasks.

When generating analytical reports, DAgent integrates the retrieved data to produce comprehensive and logically clear reports. On the DA-BIRD dataset, DAgent achieves a report accuracy score of 6.43, clearly outperforming methods like LLM-SQL (4.16) and C3 (4.76). This indicates that DAgent not only retrieves relevant data accurately but also effectively converts this data into actionable analytical conclusions.
The reports generated by DAgent are highly relevant to the query, ensuring that the information in the report is complete and closely aligned with the user's requirements. On both the DA-CCKS and DA-BIRD datasets, DAgent’s report relevance scores are higher than those of other methods, reaching 6.96 and 7.01, respectively. The experiments show that DAgent avoids redundancy and irrelevant information in the generated reports, ensuring both semantic precision and completeness.

\subsection{Efficient of Planning Strategies}
\vspace{0.4mm}
In this section, we explore the impact of problem decomposition planning, data retrieval planning, and SQL rewrite planning on the overall system efficiency through ablation experiments. The experiments validate the improvement in problem accuracy by comparing the performance differences between using and not using each planning strategy.

\vspace{0.64mm}
\textbf{Problem Decomposition Planning}
The goal of problem decomposition planning is to break down complex user question into multiple sub-questions, thereby improving the system's retrieval accuracy. In the experiments, we compared the performance differences between using and not using the problem decomposition tool. As shown in Figure \ref{fig:ab}(a) and Figure \ref{fig:ab}(d), the system with the problem decomposition tool exhibited significant improvement in retrieval performance on the DA-dataset. Especially when handling complex queries involving multiple knowledge points, question decomposition helped the system comprehensively retrieve relevant information from multiple tables.

\vspace{0.64mm}
\textbf{Data Retrieval Planning} 
In terms of data retrieval planning, we compared the performance of using a fixed single retrieval strategy (such as Text-to-SQL) with the performance of using a combined strategy of both Text-to-SQL and direct encoding and retrieval. The experimental results show that the system with the hybrid retrieval strategy performs the best in retrieval performance. As shown in Figure \ref{fig:ab}(b) and \ref{fig:ab}(e), when Text-to-SQL and the direct encoding and retrieval strategies are combined, the system’s Precision and Recall both show significant improvement, exceeding single retrieval strategies by several percentage points. This indicates that flexibly selecting and combining different data retrieval strategies can significantly enhance retrieval accuracy and data coverage.

\textbf{SQL Rewrite Planning}
The main goal of SQL rewrite planning is to optimize the generated SQL queries and reduce large scale data. The experimental results in Figure \ref{fig:ab}(c) and \ref{fig:ab}(f) show that, by comparing the execution results of SQL queries before and after optimization, the retrieval performance did not significantly decrease with the optimized SQL queries. This is because SQL rewrite primarily optimizes the query by introducing such as aggregation operations, rather than directly retrieving partial or complete data, thereby enhancing query efficiency.

\section{Conclusions}

In this paper, we introduced DAgent, an LLM-based agent system specifically designed for relational database-driven data analysis report generation. Through key modules such as task decomposition, data retrieval, and SQL optimization, DAgent significantly enhances retrieval accuracy and report generation quality in complex analytical query scenarios. Experimental results demonstrate that DAgent outperforms existing baseline methods across multiple evaluation metrics. 
The novelty of DAgent lies in its modular design, which allows the system to efficiently decompose complex tasks, flexibly retrieve relevant data, and generate reports that meet user requirements. Particularly in fields such as financial analysis and healthcare, DAgent has shown strong practical potential, effectively handling large-scale data analysis tasks and delivering accurate, complete, and concise analytical reports.  
Future work will focus on further optimizing DAgent’s submodules such as task decomposition and data retrieval, and exploring its application in more complex analytical scenarios. Additionally, enhancements in integrated reasoning mechanisms will be pursued to elevate its capabilities in relational database-driven data analysis.


\bibliographystyle{ACM-Reference-Format}
\bibliography{ref}


\begin{thebibliography}{46}


\ifx \showCODEN    \undefined \def \showCODEN     #1{\unskip}     \fi
\ifx \showDOI      \undefined \def \showDOI       #1{#1}\fi
\ifx \showISBNx    \undefined \def \showISBNx     #1{\unskip}     \fi
\ifx \showISBNxiii \undefined \def \showISBNxiii  #1{\unskip}     \fi
\ifx \showISSN     \undefined \def \showISSN      #1{\unskip}     \fi
\ifx \showLCCN     \undefined \def \showLCCN      #1{\unskip}     \fi
\ifx \shownote     \undefined \def \shownote      #1{#1}          \fi
\ifx \showarticletitle \undefined \def \showarticletitle #1{#1}   \fi
\ifx \showURL      \undefined \def \showURL       {\relax}        \fi
\providecommand\bibfield[2]{#2}
\providecommand\bibinfo[2]{#2}
\providecommand\natexlab[1]{#1}
\providecommand\showeprint[2][]{arXiv:#2}

\bibitem[\protect\citeauthoryear{{Cao}, {Chen}, {Chen}, {Zhao}, {Zhu}, and {Yu}}{{Cao} et~al\mbox{.}}{2021}]%
        {2021arXiv210601093C}
\bibfield{author}{\bibinfo{person}{Ruisheng {Cao}}, \bibinfo{person}{Lu {Chen}}, \bibinfo{person}{Zhi {Chen}}, \bibinfo{person}{Yanbin {Zhao}}, \bibinfo{person}{Su {Zhu}}, {and} \bibinfo{person}{Kai {Yu}}.} \bibinfo{year}{2021}\natexlab{}.
\newblock \showarticletitle{{LGESQL: Line Graph Enhanced Text-to-SQL Model with Mixed Local and Non-Local Relations}}.
\newblock \bibinfo{journal}{\emph{arXiv preprint arXiv:2106.01093}} (\bibinfo{year}{2021}).
\newblock


\bibitem[\protect\citeauthoryear{Chen, Miculicich, Eisenschlos, Wang, Wang, Chen, Fujii, Lin, Lee, and Pfister}{Chen et~al\mbox{.}}{2024}]%
        {chen2024tablerag}
\bibfield{author}{\bibinfo{person}{Si-An Chen}, \bibinfo{person}{Lesly Miculicich}, \bibinfo{person}{Julian~Martin Eisenschlos}, \bibinfo{person}{Zifeng Wang}, \bibinfo{person}{Zilong Wang}, \bibinfo{person}{Yanfei Chen}, \bibinfo{person}{Yasuhisa Fujii}, \bibinfo{person}{Hsuan-Tien Lin}, \bibinfo{person}{Chen-Yu Lee}, {and} \bibinfo{person}{Tomas Pfister}.} \bibinfo{year}{2024}\natexlab{}.
\newblock \showarticletitle{Tablerag: Million-token table understanding with language models}.
\newblock \bibinfo{journal}{\emph{arXiv preprint arXiv:2410.04739}} (\bibinfo{year}{2024}).
\newblock


\bibitem[\protect\citeauthoryear{{Chen}, {Chang}, {Schlinger}, {Wang}, and {Cohen}}{{Chen} et~al\mbox{.}}{2020}]%
        {ottqa}
\bibfield{author}{\bibinfo{person}{Wenhu {Chen}}, \bibinfo{person}{Ming-Wei {Chang}}, \bibinfo{person}{Eva {Schlinger}}, \bibinfo{person}{William {Wang}}, {and} \bibinfo{person}{William~W. {Cohen}}.} \bibinfo{year}{2020}\natexlab{}.
\newblock \showarticletitle{{Open Question Answering over Tables and Text}}.
\newblock \bibinfo{journal}{\emph{arXiv preprint arXiv:2010.10439}} (\bibinfo{year}{2020}).
\newblock


\bibitem[\protect\citeauthoryear{{Chen}, {Chen}, {Smiley}, {Shah}, {Borova}, {Langdon}, {Moussa}, {Beane}, {Huang}, {Routledge}, and {Wang}}{{Chen} et~al\mbox{.}}{2021}]%
        {2021arXiv210900122C}
\bibfield{author}{\bibinfo{person}{Zhiyu {Chen}}, \bibinfo{person}{Wenhu {Chen}}, \bibinfo{person}{Charese {Smiley}}, \bibinfo{person}{Sameena {Shah}}, \bibinfo{person}{Iana {Borova}}, \bibinfo{person}{Dylan {Langdon}}, \bibinfo{person}{Reema {Moussa}}, \bibinfo{person}{Matt {Beane}}, \bibinfo{person}{Ting-Hao {Huang}}, \bibinfo{person}{Bryan {Routledge}}, {and} \bibinfo{person}{William~Yang {Wang}}.} \bibinfo{year}{2021}\natexlab{}.
\newblock \showarticletitle{{FinQA: A Dataset of Numerical Reasoning over Financial Data}}.
\newblock \bibinfo{journal}{\emph{arXiv preprint arXiv:2109.00122}} (\bibinfo{year}{2021}).
\newblock


\bibitem[\protect\citeauthoryear{Deng, Chen, and Zhang}{Deng et~al\mbox{.}}{2022}]%
        {deng-etal-2022-recent}
\bibfield{author}{\bibinfo{person}{Naihao Deng}, \bibinfo{person}{Yulong Chen}, {and} \bibinfo{person}{Yue Zhang}.} \bibinfo{year}{2022}\natexlab{}.
\newblock \showarticletitle{Recent Advances in Text-to-{SQL}: A Survey of What We Have and What We Expect}. In \bibinfo{booktitle}{\emph{ICML}}.
\newblock


\bibitem[\protect\citeauthoryear{Dong, Zhang, Ge, Mao, Gao, Lin, Lou, et~al\mbox{.}}{Dong et~al\mbox{.}}{2023}]%
        {dong2023c3}
\bibfield{author}{\bibinfo{person}{Xuemei Dong}, \bibinfo{person}{Chao Zhang}, \bibinfo{person}{Yuhang Ge}, \bibinfo{person}{Yuren Mao}, \bibinfo{person}{Yunjun Gao}, \bibinfo{person}{Jinshu Lin}, \bibinfo{person}{Dongfang Lou}, {et~al\mbox{.}}} \bibinfo{year}{2023}\natexlab{}.
\newblock \showarticletitle{C3: Zero-shot text-to-sql with chatgpt}.
\newblock \bibinfo{journal}{\emph{arXiv preprint arXiv:2307.07306}} (\bibinfo{year}{2023}).
\newblock


\bibitem[\protect\citeauthoryear{{Dou}, {Gao}, {Liu}, {Pan}, {Wang}, {Che}, {Zhan}, {Kan}, and {Lou}}{{Dou} et~al\mbox{.}}{2023}]%
        {2023arXiv230101067D}
\bibfield{author}{\bibinfo{person}{Longxu {Dou}}, \bibinfo{person}{Yan {Gao}}, \bibinfo{person}{Xuqi {Liu}}, \bibinfo{person}{Mingyang {Pan}}, \bibinfo{person}{Dingzirui {Wang}}, \bibinfo{person}{Wanxiang {Che}}, \bibinfo{person}{Dechen {Zhan}}, \bibinfo{person}{Min-Yen {Kan}}, {and} \bibinfo{person}{Jian-Guang {Lou}}.} \bibinfo{year}{2023}\natexlab{}.
\newblock \showarticletitle{{Towards Knowledge-Intensive Text-to-SQL Semantic Parsing with Formulaic Knowledge}}.
\newblock \bibinfo{journal}{\emph{arXiv preprint arXiv:2301.01067}} (\bibinfo{year}{2023}).
\newblock


\bibitem[\protect\citeauthoryear{{Eisenschlos}, {Gor}, {M{\"u}ller}, and {Cohen}}{{Eisenschlos} et~al\mbox{.}}{2021}]%
        {2021arXiv210904312E}
\bibfield{author}{\bibinfo{person}{Julian~Martin {Eisenschlos}}, \bibinfo{person}{Maharshi {Gor}}, \bibinfo{person}{Thomas {M{\"u}ller}}, {and} \bibinfo{person}{William~W. {Cohen}}.} \bibinfo{year}{2021}\natexlab{}.
\newblock \showarticletitle{{MATE: Multi-view Attention for Table Transformer Efficiency}}.
\newblock \bibinfo{journal}{\emph{arXiv preprint arXiv:2109.04312}} (\bibinfo{year}{2021}).
\newblock


\bibitem[\protect\citeauthoryear{Es, James, Espinosa-Anke, and Schockaert}{Es et~al\mbox{.}}{2023}]%
        {es2023ragas}
\bibfield{author}{\bibinfo{person}{Shahul Es}, \bibinfo{person}{Jithin James}, \bibinfo{person}{Luis Espinosa-Anke}, {and} \bibinfo{person}{Steven Schockaert}.} \bibinfo{year}{2023}\natexlab{}.
\newblock \showarticletitle{Ragas: Automated evaluation of retrieval augmented generation}.
\newblock \bibinfo{journal}{\emph{arXiv preprint arXiv:2309.15217}} (\bibinfo{year}{2023}).
\newblock


\bibitem[\protect\citeauthoryear{{Gan}, {Chen}, {Huang}, and {Purver}}{{Gan} et~al\mbox{.}}{2022}]%
        {spider-ss-cg}
\bibfield{author}{\bibinfo{person}{Yujian {Gan}}, \bibinfo{person}{Xinyun {Chen}}, \bibinfo{person}{Qiuping {Huang}}, {and} \bibinfo{person}{Matthew {Purver}}.} \bibinfo{year}{2022}\natexlab{}.
\newblock \showarticletitle{{Measuring and Improving Compositional Generalization in Text-to-SQL via Component Alignment}}.
\newblock \bibinfo{journal}{\emph{arXiv preprint arXiv:2205.02054}} (\bibinfo{year}{2022}).
\newblock


\bibitem[\protect\citeauthoryear{{Gao}, {Wang}, {Li}, {Sun}, {Qian}, {Ding}, and {Zhou}}{{Gao} et~al\mbox{.}}{2023}]%
        {2023arXiv230815363G}
\bibfield{author}{\bibinfo{person}{Dawei {Gao}}, \bibinfo{person}{Haibin {Wang}}, \bibinfo{person}{Yaliang {Li}}, \bibinfo{person}{Xiuyu {Sun}}, \bibinfo{person}{Yichen {Qian}}, \bibinfo{person}{Bolin {Ding}}, {and} \bibinfo{person}{Jingren {Zhou}}.} \bibinfo{year}{2023}\natexlab{}.
\newblock \showarticletitle{{Text-to-SQL Empowered by Large Language Models: A Benchmark Evaluation}}.
\newblock \bibinfo{journal}{\emph{arXiv preprint arXiv:2308.15363}} (\bibinfo{year}{2023}).
\newblock


\bibitem[\protect\citeauthoryear{{Guo}, {Zhan}, {Gao}, {Xiao}, {Lou}, {Liu}, and {Zhang}}{{Guo} et~al\mbox{.}}{2019}]%
        {2019arXiv190508205G}
\bibfield{author}{\bibinfo{person}{Jiaqi {Guo}}, \bibinfo{person}{Zecheng {Zhan}}, \bibinfo{person}{Yan {Gao}}, \bibinfo{person}{Yan {Xiao}}, \bibinfo{person}{Jian-Guang {Lou}}, \bibinfo{person}{Ting {Liu}}, {and} \bibinfo{person}{Dongmei {Zhang}}.} \bibinfo{year}{2019}\natexlab{}.
\newblock \showarticletitle{{Towards Complex Text-to-SQL in Cross-Domain Database with Intermediate Representation}}.
\newblock \bibinfo{journal}{\emph{arXiv preprint arXiv:1905.08205}} (\bibinfo{year}{2019}).
\newblock


\bibitem[\protect\citeauthoryear{Guo, Tahir, Zhang, Wang, Ward, et~al\mbox{.}}{Guo et~al\mbox{.}}{2024b}]%
        {guo2024automatic}
\bibfield{author}{\bibinfo{person}{Li Guo}, \bibinfo{person}{Anas~M Tahir}, \bibinfo{person}{Dong Zhang}, \bibinfo{person}{Z~Jane Wang}, \bibinfo{person}{Rabab~K Ward}, {et~al\mbox{.}}} \bibinfo{year}{2024}\natexlab{b}.
\newblock \showarticletitle{Automatic Medical Report Generation: Methods and Applications}.
\newblock \bibinfo{journal}{\emph{APSIPA Transactions on Signal and Information Processing}} \bibinfo{volume}{13}, \bibinfo{number}{1} (\bibinfo{year}{2024}).
\newblock


\bibitem[\protect\citeauthoryear{Guo, Chen, Wang, Chang, Pei, Chawla, Wiest, and Zhang}{Guo et~al\mbox{.}}{2024a}]%
        {guo2024large}
\bibfield{author}{\bibinfo{person}{Taicheng Guo}, \bibinfo{person}{Xiuying Chen}, \bibinfo{person}{Yaqi Wang}, \bibinfo{person}{Ruidi Chang}, \bibinfo{person}{Shichao Pei}, \bibinfo{person}{Nitesh~V Chawla}, \bibinfo{person}{Olaf Wiest}, {and} \bibinfo{person}{Xiangliang Zhang}.} \bibinfo{year}{2024}\natexlab{a}.
\newblock \showarticletitle{Large language model based multi-agents: A survey of progress and challenges}.
\newblock \bibinfo{journal}{\emph{arXiv preprint arXiv:2402.01680}} (\bibinfo{year}{2024}).
\newblock


\bibitem[\protect\citeauthoryear{{Herzig}, {M{\"u}ller}, {Krichene}, and {Eisenschlos}}{{Herzig} et~al\mbox{.}}{2021}]%
        {nq-tables}
\bibfield{author}{\bibinfo{person}{Jonathan {Herzig}}, \bibinfo{person}{Thomas {M{\"u}ller}}, \bibinfo{person}{Syrine {Krichene}}, {and} \bibinfo{person}{Julian~Martin {Eisenschlos}}.} \bibinfo{year}{2021}\natexlab{}.
\newblock \showarticletitle{{Open Domain Question Answering over Tables via Dense Retrieval}}.
\newblock \bibinfo{journal}{\emph{arXiv preprint arXiv:2103.12011}} (\bibinfo{year}{2021}).
\newblock


\bibitem[\protect\citeauthoryear{Herzig, Nowak, M{\"u}ller, Piccinno, and Eisenschlos}{Herzig et~al\mbox{.}}{2020}]%
        {herzig2020tapas}
\bibfield{author}{\bibinfo{person}{Jonathan Herzig}, \bibinfo{person}{Pawe{\l}~Krzysztof Nowak}, \bibinfo{person}{Thomas M{\"u}ller}, \bibinfo{person}{Francesco Piccinno}, {and} \bibinfo{person}{Julian~Martin Eisenschlos}.} \bibinfo{year}{2020}\natexlab{}.
\newblock \showarticletitle{TaPas: Weakly supervised table parsing via pre-training}.
\newblock \bibinfo{journal}{\emph{arXiv preprint arXiv:2004.02349}} (\bibinfo{year}{2020}).
\newblock


\bibitem[\protect\citeauthoryear{Hong, Yuan, Zhang, Chen, Dong, Huang, and Huang}{Hong et~al\mbox{.}}{2024}]%
        {hong2024next}
\bibfield{author}{\bibinfo{person}{Zijin Hong}, \bibinfo{person}{Zheng Yuan}, \bibinfo{person}{Qinggang Zhang}, \bibinfo{person}{Hao Chen}, \bibinfo{person}{Junnan Dong}, \bibinfo{person}{Feiran Huang}, {and} \bibinfo{person}{Xiao Huang}.} \bibinfo{year}{2024}\natexlab{}.
\newblock \showarticletitle{Next-Generation Database Interfaces: A Survey of LLM-based Text-to-SQL}.
\newblock \bibinfo{journal}{\emph{arXiv preprint arXiv:2406.08426}} (\bibinfo{year}{2024}).
\newblock


\bibitem[\protect\citeauthoryear{Hu, Shen, Wallis, Allen-Zhu, Li, Wang, Wang, and Chen}{Hu et~al\mbox{.}}{2021}]%
        {hu2021lora}
\bibfield{author}{\bibinfo{person}{Edward~J Hu}, \bibinfo{person}{Yelong Shen}, \bibinfo{person}{Phillip Wallis}, \bibinfo{person}{Zeyuan Allen-Zhu}, \bibinfo{person}{Yuanzhi Li}, \bibinfo{person}{Shean Wang}, \bibinfo{person}{Lu Wang}, {and} \bibinfo{person}{Weizhu Chen}.} \bibinfo{year}{2021}\natexlab{}.
\newblock \showarticletitle{Lora: Low-rank adaptation of large language models}.
\newblock \bibinfo{journal}{\emph{arXiv preprint arXiv:2106.09685}} (\bibinfo{year}{2021}).
\newblock


\bibitem[\protect\citeauthoryear{{Hui}, {Geng}, {Wang}, {Qin}, {Li}, {Sun}, and {Li}}{{Hui} et~al\mbox{.}}{2022}]%
        {2022arXiv220306958H}
\bibfield{author}{\bibinfo{person}{Binyuan {Hui}}, \bibinfo{person}{Ruiying {Geng}}, \bibinfo{person}{Lihan {Wang}}, \bibinfo{person}{Bowen {Qin}}, \bibinfo{person}{Bowen {Li}}, \bibinfo{person}{Jian {Sun}}, {and} \bibinfo{person}{Yongbin {Li}}.} \bibinfo{year}{2022}\natexlab{}.
\newblock \showarticletitle{{S$^2$SQL: Injecting Syntax to Question-Schema Interaction Graph Encoder for Text-to-SQL Parsers}}.
\newblock \bibinfo{journal}{\emph{arXiv preprint arXiv:2203.06958}} (\bibinfo{year}{2022}).
\newblock


\bibitem[\protect\citeauthoryear{Iyyer, Yih, and Chang}{Iyyer et~al\mbox{.}}{2017}]%
        {sqa}
\bibfield{author}{\bibinfo{person}{Mohit Iyyer}, \bibinfo{person}{Wen-tau Yih}, {and} \bibinfo{person}{Ming-Wei Chang}.} \bibinfo{year}{2017}\natexlab{}.
\newblock \showarticletitle{Search-based Neural Structured Learning for Sequential Question Answering}. In \bibinfo{booktitle}{\emph{ACL}}.
\newblock


\bibitem[\protect\citeauthoryear{Jin, Siebert, Li, and Chen}{Jin et~al\mbox{.}}{2022}]%
        {jin2022survey}
\bibfield{author}{\bibinfo{person}{Nengzheng Jin}, \bibinfo{person}{Joanna Siebert}, \bibinfo{person}{Dongfang Li}, {and} \bibinfo{person}{Qingcai Chen}.} \bibinfo{year}{2022}\natexlab{}.
\newblock \showarticletitle{A survey on table question answering: recent advances}. In \bibinfo{booktitle}{\emph{China Conference on Knowledge Graph and Semantic Computing}}. Springer, \bibinfo{pages}{174--186}.
\newblock


\bibitem[\protect\citeauthoryear{{Li}, {Hui}, {Qu}, {Yang}, {Li}, {Li}, {Wang}, {Qin}, {Cao}, {Geng}, {Huo}, {Zhou}, {Ma}, {Li}, {Chang}, {Huang}, {Cheng}, and {Li}}{{Li} et~al\mbox{.}}{2023a}]%
        {2023arXiv230503111L}
\bibfield{author}{\bibinfo{person}{Jinyang {Li}}, \bibinfo{person}{Binyuan {Hui}}, \bibinfo{person}{Ge {Qu}}, \bibinfo{person}{Jiaxi {Yang}}, \bibinfo{person}{Binhua {Li}}, \bibinfo{person}{Bowen {Li}}, \bibinfo{person}{Bailin {Wang}}, \bibinfo{person}{Bowen {Qin}}, \bibinfo{person}{Rongyu {Cao}}, \bibinfo{person}{Ruiying {Geng}}, \bibinfo{person}{Nan {Huo}}, \bibinfo{person}{Xuanhe {Zhou}}, \bibinfo{person}{Chenhao {Ma}}, \bibinfo{person}{Guoliang {Li}}, \bibinfo{person}{Kevin C.~C. {Chang}}, \bibinfo{person}{Fei {Huang}}, \bibinfo{person}{Reynold {Cheng}}, {and} \bibinfo{person}{Yongbin {Li}}.} \bibinfo{year}{2023}\natexlab{a}.
\newblock \showarticletitle{{Can LLM Already Serve as A Database Interface? A BIg Bench for Large-Scale Database Grounded Text-to-SQLs}}.
\newblock \bibinfo{journal}{\emph{arXiv preprint arXiv:2305.03111}} (\bibinfo{year}{2023}).
\newblock


\bibitem[\protect\citeauthoryear{{Li}, {Hui}, {Qu}, {Yang}, {Li}, {Li}, {Wang}, {Qin}, {Cao}, {Geng}, {Huo}, {Zhou}, {Ma}, {Li}, {Chang}, {Huang}, {Cheng}, and {Li}}{{Li} et~al\mbox{.}}{2023b}]%
        {2023bird}
\bibfield{author}{\bibinfo{person}{Jinyang {Li}}, \bibinfo{person}{Binyuan {Hui}}, \bibinfo{person}{Ge {Qu}}, \bibinfo{person}{Jiaxi {Yang}}, \bibinfo{person}{Binhua {Li}}, \bibinfo{person}{Bowen {Li}}, \bibinfo{person}{Bailin {Wang}}, \bibinfo{person}{Bowen {Qin}}, \bibinfo{person}{Rongyu {Cao}}, \bibinfo{person}{Ruiying {Geng}}, \bibinfo{person}{Nan {Huo}}, \bibinfo{person}{Xuanhe {Zhou}}, \bibinfo{person}{Chenhao {Ma}}, \bibinfo{person}{Guoliang {Li}}, \bibinfo{person}{Kevin C.~C. {Chang}}, \bibinfo{person}{Fei {Huang}}, \bibinfo{person}{Reynold {Cheng}}, {and} \bibinfo{person}{Yongbin {Li}}.} \bibinfo{year}{2023}\natexlab{b}.
\newblock \showarticletitle{{Can LLM Already Serve as A Database Interface? A BIg Bench for Large-Scale Database Grounded Text-to-SQLs}}.
\newblock \bibinfo{journal}{\emph{arXiv preprint arXiv:2305.03111}} (\bibinfo{year}{2023}).
\newblock


\bibitem[\protect\citeauthoryear{Lu, Zhang, Fan, Fu, Chen, and Du}{Lu et~al\mbox{.}}{2024}]%
        {lu2024large}
\bibfield{author}{\bibinfo{person}{Weizheng Lu}, \bibinfo{person}{Jing Zhang}, \bibinfo{person}{Ju Fan}, \bibinfo{person}{Zihao Fu}, \bibinfo{person}{Yueguo Chen}, {and} \bibinfo{person}{Xiaoyong Du}.} \bibinfo{year}{2024}\natexlab{}.
\newblock \showarticletitle{Large language model for table processing: A survey}.
\newblock \bibinfo{journal}{\emph{arXiv preprint arXiv:2402.05121}} (\bibinfo{year}{2024}).
\newblock


\bibitem[\protect\citeauthoryear{Mao, Ge, Fan, Xu, Mi, Hu, and Gao}{Mao et~al\mbox{.}}{2025}]%
        {mao2025survey}
\bibfield{author}{\bibinfo{person}{Yuren Mao}, \bibinfo{person}{Yuhang Ge}, \bibinfo{person}{Yijiang Fan}, \bibinfo{person}{Wenyi Xu}, \bibinfo{person}{Yu Mi}, \bibinfo{person}{Zhonghao Hu}, {and} \bibinfo{person}{Yunjun Gao}.} \bibinfo{year}{2025}\natexlab{}.
\newblock \showarticletitle{A survey on lora of large language models}.
\newblock \bibinfo{journal}{\emph{Frontiers of Computer Science}} \bibinfo{volume}{19}, \bibinfo{number}{7} (\bibinfo{year}{2025}), \bibinfo{pages}{197605}.
\newblock


\bibitem[\protect\citeauthoryear{{Nan}, {Hsieh}, {Mao}, {Lin}, {Verma}, {Zhang}, {Kry{\'s}ci{\'n}ski}, {Schoelkopf}, {Kong}, {Tang}, {Mutuma}, {Rosand}, {Trindade}, {Bandaru}, {Cunningham}, {Xiong}, and {Radev}}{{Nan} et~al\mbox{.}}{2021}]%
        {2021arXiv210400369N}
\bibfield{author}{\bibinfo{person}{Linyong {Nan}}, \bibinfo{person}{Chiachun {Hsieh}}, \bibinfo{person}{Ziming {Mao}}, \bibinfo{person}{Xi~Victoria {Lin}}, \bibinfo{person}{Neha {Verma}}, \bibinfo{person}{Rui {Zhang}}, \bibinfo{person}{Wojciech {Kry{\'s}ci{\'n}ski}}, \bibinfo{person}{Nick {Schoelkopf}}, \bibinfo{person}{Riley {Kong}}, \bibinfo{person}{Xiangru {Tang}}, \bibinfo{person}{Murori {Mutuma}}, \bibinfo{person}{Ben {Rosand}}, \bibinfo{person}{Isabel {Trindade}}, \bibinfo{person}{Renusree {Bandaru}}, \bibinfo{person}{Jacob {Cunningham}}, \bibinfo{person}{Caiming {Xiong}}, {and} \bibinfo{person}{Dragomir {Radev}}.} \bibinfo{year}{2021}\natexlab{}.
\newblock \showarticletitle{{FeTaQA: Free-form Table Question Answering}}.
\newblock \bibinfo{journal}{\emph{arXiv preprint arXiv:2104.00369}} (\bibinfo{year}{2021}).
\newblock


\bibitem[\protect\citeauthoryear{{Pasupat} and {Liang}}{{Pasupat} and {Liang}}{2015}]%
        {wtq}
\bibfield{author}{\bibinfo{person}{Panupong {Pasupat}} {and} \bibinfo{person}{Percy {Liang}}.} \bibinfo{year}{2015}\natexlab{}.
\newblock \showarticletitle{{Compositional Semantic Parsing on Semi-Structured Tables}}.
\newblock \bibinfo{journal}{\emph{arXiv preprint arXiv:1508.00305}} (\bibinfo{year}{2015}).
\newblock


\bibitem[\protect\citeauthoryear{{Pi}, {Wang}, {Gao}, {Guo}, {Li}, and {Lou}}{{Pi} et~al\mbox{.}}{2022}]%
        {adveta}
\bibfield{author}{\bibinfo{person}{Xinyu {Pi}}, \bibinfo{person}{Bing {Wang}}, \bibinfo{person}{Yan {Gao}}, \bibinfo{person}{Jiaqi {Guo}}, \bibinfo{person}{Zhoujun {Li}}, {and} \bibinfo{person}{Jian-Guang {Lou}}.} \bibinfo{year}{2022}\natexlab{}.
\newblock \showarticletitle{{Towards Robustness of Text-to-SQL Models Against Natural and Realistic Adversarial Table Perturbation}}.
\newblock \bibinfo{journal}{\emph{arXiv preprint arXiv:2212.09994}} (\bibinfo{year}{2022}).
\newblock


\bibitem[\protect\citeauthoryear{{Pourreza} and {Rafiei}}{{Pourreza} and {Rafiei}}{2023}]%
        {2023arXiv230411015P}
\bibfield{author}{\bibinfo{person}{Mohammadreza {Pourreza}} {and} \bibinfo{person}{Davood {Rafiei}}.} \bibinfo{year}{2023}\natexlab{}.
\newblock \showarticletitle{{DIN-SQL: Decomposed In-Context Learning of Text-to-SQL with Self-Correction}}.
\newblock \bibinfo{journal}{\emph{arXiv preprint arXiv:2304.11015}} (\bibinfo{year}{2023}).
\newblock


\bibitem[\protect\citeauthoryear{Ravi, Amrouni, Stefanucci, Nourbakhsh, Reddy, and Veloso}{Ravi et~al\mbox{.}}{2020}]%
        {ravi2020docubot}
\bibfield{author}{\bibinfo{person}{Vineeth Ravi}, \bibinfo{person}{Selim Amrouni}, \bibinfo{person}{Andrea Stefanucci}, \bibinfo{person}{Armineh Nourbakhsh}, \bibinfo{person}{Prashant Reddy}, {and} \bibinfo{person}{Manuela Veloso}.} \bibinfo{year}{2020}\natexlab{}.
\newblock \showarticletitle{DocuBot: Generating financial reports using natural language interactions}.
\newblock \bibinfo{journal}{\emph{arXiv preprint arXiv:2010.01169}} (\bibinfo{year}{2020}).
\newblock


\bibitem[\protect\citeauthoryear{Roychowdhury, Krema, Mahammad, Moore, Mukherjee, and Prakashchandra}{Roychowdhury et~al\mbox{.}}{2024}]%
        {roychowdhury2024eratta}
\bibfield{author}{\bibinfo{person}{Sohini Roychowdhury}, \bibinfo{person}{Marko Krema}, \bibinfo{person}{Anvar Mahammad}, \bibinfo{person}{Brian Moore}, \bibinfo{person}{Arijit Mukherjee}, {and} \bibinfo{person}{Punit Prakashchandra}.} \bibinfo{year}{2024}\natexlab{}.
\newblock \showarticletitle{ERATTA: Extreme RAG for Table To Answers with Large Language Models}.
\newblock \bibinfo{journal}{\emph{arXiv preprint arXiv:2405.03963}} (\bibinfo{year}{2024}).
\newblock


\bibitem[\protect\citeauthoryear{{Shi}, {Tang}, {Zhang}, {Zhang}, and {Yang}}{{Shi} et~al\mbox{.}}{2024}]%
        {2024arXiv240715186S}
\bibfield{author}{\bibinfo{person}{Liang {Shi}}, \bibinfo{person}{Zhengju {Tang}}, \bibinfo{person}{Nan {Zhang}}, \bibinfo{person}{Xiaotong {Zhang}}, {and} \bibinfo{person}{Zhi {Yang}}.} \bibinfo{year}{2024}\natexlab{}.
\newblock \showarticletitle{{A Survey on Employing Large Language Models for Text-to-SQL Tasks}}.
\newblock \bibinfo{journal}{\emph{arXiv preprint arXiv:2407.15186}} (\bibinfo{year}{2024}).
\newblock


\bibitem[\protect\citeauthoryear{{Wang}, {Shin}, {Liu}, {Polozov}, and {Richardson}}{{Wang} et~al\mbox{.}}{2019}]%
        {2019arXiv191104942W}
\bibfield{author}{\bibinfo{person}{Bailin {Wang}}, \bibinfo{person}{Richard {Shin}}, \bibinfo{person}{Xiaodong {Liu}}, \bibinfo{person}{Oleksandr {Polozov}}, {and} \bibinfo{person}{Matthew {Richardson}}.} \bibinfo{year}{2019}\natexlab{}.
\newblock \showarticletitle{{RAT-SQL: Relation-Aware Schema Encoding and Linking for Text-to-SQL Parsers}}.
\newblock \bibinfo{journal}{\emph{arXiv preprint arXiv:1911.04942}} (\bibinfo{year}{2019}).
\newblock


\bibitem[\protect\citeauthoryear{{Wei}, {Wang}, {Schuurmans}, {Bosma}, {Ichter}, {Xia}, {Chi}, {Le}, and {Zhou}}{{Wei} et~al\mbox{.}}{2022}]%
        {2022arXiv220111903W}
\bibfield{author}{\bibinfo{person}{Jason {Wei}}, \bibinfo{person}{Xuezhi {Wang}}, \bibinfo{person}{Dale {Schuurmans}}, \bibinfo{person}{Maarten {Bosma}}, \bibinfo{person}{Brian {Ichter}}, \bibinfo{person}{Fei {Xia}}, \bibinfo{person}{Ed {Chi}}, \bibinfo{person}{Quoc {Le}}, {and} \bibinfo{person}{Denny {Zhou}}.} \bibinfo{year}{2022}\natexlab{}.
\newblock \showarticletitle{{Chain-of-Thought Prompting Elicits Reasoning in Large Language Models}}.
\newblock \bibinfo{journal}{\emph{arXiv preprint arXiv:2201.11903}} (\bibinfo{year}{2022}).
\newblock


\bibitem[\protect\citeauthoryear{Xi, Chen, Guo, He, Ding, Hong, Zhang, Wang, Jin, Zhou, et~al\mbox{.}}{Xi et~al\mbox{.}}{2023}]%
        {xi2023rise}
\bibfield{author}{\bibinfo{person}{Zhiheng Xi}, \bibinfo{person}{Wenxiang Chen}, \bibinfo{person}{Xin Guo}, \bibinfo{person}{Wei He}, \bibinfo{person}{Yiwen Ding}, \bibinfo{person}{Boyang Hong}, \bibinfo{person}{Ming Zhang}, \bibinfo{person}{Junzhe Wang}, \bibinfo{person}{Senjie Jin}, \bibinfo{person}{Enyu Zhou}, {et~al\mbox{.}}} \bibinfo{year}{2023}\natexlab{}.
\newblock \showarticletitle{The rise and potential of large language model based agents: A survey}.
\newblock \bibinfo{journal}{\emph{arXiv preprint arXiv:2309.07864}} (\bibinfo{year}{2023}).
\newblock


\bibitem[\protect\citeauthoryear{Xu, Liu, and Song}{Xu et~al\mbox{.}}{2017}]%
        {xu2017sqlnet}
\bibfield{author}{\bibinfo{person}{Xiaojun Xu}, \bibinfo{person}{Chang Liu}, {and} \bibinfo{person}{Dawn Song}.} \bibinfo{year}{2017}\natexlab{}.
\newblock \showarticletitle{Sqlnet: Generating structured queries from natural language without reinforcement learning}.
\newblock \bibinfo{journal}{\emph{arXiv preprint arXiv:1711.04436}} (\bibinfo{year}{2017}).
\newblock


\bibitem[\protect\citeauthoryear{{Yao}, {Yu}, {Zhao}, {Shafran}, {Griffiths}, {Cao}, and {Narasimhan}}{{Yao} et~al\mbox{.}}{2023}]%
        {2023arXiv230510601Y}
\bibfield{author}{\bibinfo{person}{Shunyu {Yao}}, \bibinfo{person}{Dian {Yu}}, \bibinfo{person}{Jeffrey {Zhao}}, \bibinfo{person}{Izhak {Shafran}}, \bibinfo{person}{Thomas~L. {Griffiths}}, \bibinfo{person}{Yuan {Cao}}, {and} \bibinfo{person}{Karthik {Narasimhan}}.} \bibinfo{year}{2023}\natexlab{}.
\newblock \showarticletitle{{Tree of Thoughts: Deliberate Problem Solving with Large Language Models}}.
\newblock \bibinfo{journal}{\emph{arXiv preprint arXiv:2305.10601}} (\bibinfo{year}{2023}).
\newblock


\bibitem[\protect\citeauthoryear{Yao, Zhao, Yu, Du, Shafran, Narasimhan, and Cao}{Yao et~al\mbox{.}}{2022}]%
        {yao2022react}
\bibfield{author}{\bibinfo{person}{Shunyu Yao}, \bibinfo{person}{Jeffrey Zhao}, \bibinfo{person}{Dian Yu}, \bibinfo{person}{Nan Du}, \bibinfo{person}{Izhak Shafran}, \bibinfo{person}{Karthik Narasimhan}, {and} \bibinfo{person}{Yuan Cao}.} \bibinfo{year}{2022}\natexlab{}.
\newblock \showarticletitle{React: Synergizing reasoning and acting in language models}.
\newblock \bibinfo{journal}{\emph{arXiv preprint arXiv:2210.03629}} (\bibinfo{year}{2022}).
\newblock


\bibitem[\protect\citeauthoryear{{Yin}, {Neubig}, {Yih}, and {Riedel}}{{Yin} et~al\mbox{.}}{2020}]%
        {2020arXiv200508314Y}
\bibfield{author}{\bibinfo{person}{Pengcheng {Yin}}, \bibinfo{person}{Graham {Neubig}}, \bibinfo{person}{Wen-tau {Yih}}, {and} \bibinfo{person}{Sebastian {Riedel}}.} \bibinfo{year}{2020}\natexlab{}.
\newblock \showarticletitle{{TaBERT: Pretraining for Joint Understanding of Textual and Tabular Data}}.
\newblock \bibinfo{journal}{\emph{arXiv preprint arXiv: 2005.08314}} (\bibinfo{year}{2020}).
\newblock


\bibitem[\protect\citeauthoryear{Yu, Li, Zhang, Zhang, and Radev}{Yu et~al\mbox{.}}{2018}]%
        {yu-etal-2018-typesql}
\bibfield{author}{\bibinfo{person}{Tao Yu}, \bibinfo{person}{Zifan Li}, \bibinfo{person}{Zilin Zhang}, \bibinfo{person}{Rui Zhang}, {and} \bibinfo{person}{Dragomir Radev}.} \bibinfo{year}{2018}\natexlab{}.
\newblock \showarticletitle{{T}ype{SQL}: Knowledge-Based Type-Aware Neural Text-to-{SQL} Generation}. In \bibinfo{booktitle}{\emph{ACL}}. \bibinfo{pages}{588--594}.
\newblock


\bibitem[\protect\citeauthoryear{{Yu}, {Zhang}, {Yang}, {Yasunaga}, {Wang}, {Li}, {Ma}, {Li}, {Yao}, {Roman}, {Zhang}, and {Radev}}{{Yu} et~al\mbox{.}}{2018a}]%
        {2018arXiv180908887Y}
\bibfield{author}{\bibinfo{person}{Tao {Yu}}, \bibinfo{person}{Rui {Zhang}}, \bibinfo{person}{Kai {Yang}}, \bibinfo{person}{Michihiro {Yasunaga}}, \bibinfo{person}{Dongxu {Wang}}, \bibinfo{person}{Zifan {Li}}, \bibinfo{person}{James {Ma}}, \bibinfo{person}{Irene {Li}}, \bibinfo{person}{Qingning {Yao}}, \bibinfo{person}{Shanelle {Roman}}, \bibinfo{person}{Zilin {Zhang}}, {and} \bibinfo{person}{Dragomir {Radev}}.} \bibinfo{year}{2018}\natexlab{a}.
\newblock \showarticletitle{{Spider: A Large-Scale Human-Labeled Dataset for Complex and Cross-Domain Semantic Parsing and Text-to-SQL Task}}.
\newblock \bibinfo{journal}{\emph{arXiv preprint arXiv:1809.08887}} (\bibinfo{year}{2018}).
\newblock


\bibitem[\protect\citeauthoryear{{Yu}, {Zhang}, {Yang}, {Yasunaga}, {Wang}, {Li}, {Ma}, {Li}, {Yao}, {Roman}, {Zhang}, and {Radev}}{{Yu} et~al\mbox{.}}{2018b}]%
        {2018Spider}
\bibfield{author}{\bibinfo{person}{Tao {Yu}}, \bibinfo{person}{Rui {Zhang}}, \bibinfo{person}{Kai {Yang}}, \bibinfo{person}{Michihiro {Yasunaga}}, \bibinfo{person}{Dongxu {Wang}}, \bibinfo{person}{Zifan {Li}}, \bibinfo{person}{James {Ma}}, \bibinfo{person}{Irene {Li}}, \bibinfo{person}{Qingning {Yao}}, \bibinfo{person}{Shanelle {Roman}}, \bibinfo{person}{Zilin {Zhang}}, {and} \bibinfo{person}{Dragomir {Radev}}.} \bibinfo{year}{2018}\natexlab{b}.
\newblock \showarticletitle{{Spider: A Large-Scale Human-Labeled Dataset for Complex and Cross-Domain Semantic Parsing and Text-to-SQL Task}}.
\newblock \bibinfo{journal}{\emph{arXiv preprint arXiv:1809.08887}} (\bibinfo{year}{2018}).
\newblock


\bibitem[\protect\citeauthoryear{Zhang, Mao, Fan, Mi, Gao, Chen, Lou, and Lin}{Zhang et~al\mbox{.}}{2024}]%
        {zhang2024finsql}
\bibfield{author}{\bibinfo{person}{Chao Zhang}, \bibinfo{person}{Yuren Mao}, \bibinfo{person}{Yijiang Fan}, \bibinfo{person}{Yu Mi}, \bibinfo{person}{Yunjun Gao}, \bibinfo{person}{Lu Chen}, \bibinfo{person}{Dongfang Lou}, {and} \bibinfo{person}{Jinshu Lin}.} \bibinfo{year}{2024}\natexlab{}.
\newblock \showarticletitle{FinSQL: Model-Agnostic LLMs-based Text-to-SQL Framework for Financial Analysis}.
\newblock \bibinfo{journal}{\emph{arXiv preprint arXiv:2401.10506}} (\bibinfo{year}{2024}).
\newblock


\bibitem[\protect\citeauthoryear{{Zhang}, {Cao}, {Chen}, {Xu}, and {Yu}}{{Zhang} et~al\mbox{.}}{2023}]%
        {2023arXiv231017342Z}
\bibfield{author}{\bibinfo{person}{Hanchong {Zhang}}, \bibinfo{person}{Ruisheng {Cao}}, \bibinfo{person}{Lu {Chen}}, \bibinfo{person}{Hongshen {Xu}}, {and} \bibinfo{person}{Kai {Yu}}.} \bibinfo{year}{2023}\natexlab{}.
\newblock \showarticletitle{{ACT-SQL: In-Context Learning for Text-to-SQL with Automatically-Generated Chain-of-Thought}}.
\newblock \bibinfo{journal}{\emph{arXiv preprint arXiv:2310.17342}} (\bibinfo{year}{2023}).
\newblock


\bibitem[\protect\citeauthoryear{{Zhong}, {Xiong}, and {Socher}}{{Zhong} et~al\mbox{.}}{2017}]%
        {2017wikisql}
\bibfield{author}{\bibinfo{person}{Victor {Zhong}}, \bibinfo{person}{Caiming {Xiong}}, {and} \bibinfo{person}{Richard {Socher}}.} \bibinfo{year}{2017}\natexlab{}.
\newblock \showarticletitle{{Seq2SQL: Generating Structured Queries from Natural Language using Reinforcement Learning}}.
\newblock \bibinfo{journal}{\emph{arXiv preprint arXiv:1709.00103}} (\bibinfo{year}{2017}).
\newblock


\bibitem[\protect\citeauthoryear{{Zhu}, {Lei}, {Huang}, {Wang}, {Zhang}, {Lv}, {Feng}, and {Chua}}{{Zhu} et~al\mbox{.}}{2021}]%
        {2021arXiv210507624Z}
\bibfield{author}{\bibinfo{person}{Fengbin {Zhu}}, \bibinfo{person}{Wenqiang {Lei}}, \bibinfo{person}{Youcheng {Huang}}, \bibinfo{person}{Chao {Wang}}, \bibinfo{person}{Shuo {Zhang}}, \bibinfo{person}{Jiancheng {Lv}}, \bibinfo{person}{Fuli {Feng}}, {and} \bibinfo{person}{Tat-Seng {Chua}}.} \bibinfo{year}{2021}\natexlab{}.
\newblock \showarticletitle{{TAT-QA: A Question Answering Benchmark on a Hybrid of Tabular and Textual Content in Finance}}.
\newblock \bibinfo{journal}{\emph{arXiv preprint arXiv:2105.07624}} (\bibinfo{year}{2021}).
\newblock


\end{thebibliography}

\end{document}